\documentclass[preprint2]{aastex}

\usepackage{subfigure}
\usepackage{natbib}
\usepackage{graphicx}
\slugcomment{Draft: \today}

\shorttitle{Spitzer/IRS investigation of MIPSGAL 24~$\mu$m compact bubbles.}
\shortauthors{Flagey et al.}

\begin{document}

\title{Spitzer/IRS investigation of MIPSGAL 24 microns compact bubbles.}

\author{N. Flagey\altaffilmark{1}}
\author{A. Noriega-Crespo\altaffilmark{1}}
\author{N. Billot\altaffilmark{2}}
\author{S.J. Carey\altaffilmark{1}}

\email{nflagey@ipac.caltech.edu}

\altaffiltext{1}{Spitzer Science Center, California Institute of Technology, 1200 East California Boulevard, MC 220-6, Pasadena, CA 91125, USA}
\altaffiltext{2}{NASA Herschel Science Center, California Institute of Technology, 770 S. Wilson Ave, MS 100-22, Pasadena, CA 91125, USA}

\begin{abstract}
The MIPSGAL 24~$\mu$m Galactic Plane Survey has revealed more than 400 compact-extended objects. Less than 15\% of these MIPSGAL bubbles (MBs) are known and identified as evolved stars. We present Spitzer observations of 4 MBs obtained with the InfraRed Spectrograph to determine the origin of the mid-IR emission. We model the mid-IR gas lines and the dust emission to infer physical conditions within the MBs and consequently their nature. Two MBs show a dust-poor spectrum dominated by highly ionized gas lines of [\ion{O}{4}], [\ion{Ne}{3}], [\ion{Ne}{5}], [\ion{S}{3}] and [\ion{S}{4}]. We identify them as planetary nebulae with a density of a few 10$^3\ \rm{cm^{-3}}$ and a central white dwarf of $\gtrsim 200,000$ K. The mid-IR emission of the two other MBs is dominated by a dust continuum and lower-excitation lines. Both of them show a central source in the near-IR (2MASS and IRAC) broadband images. The first dust-rich MB matches a Wolf-Rayet star of $\sim 60,000$ K at 7.5 kpc with dust components of $\sim170$ and $\sim1750$ K. Its mass is about $10^{-3}\ \rm{M_\odot}$ and its mass loss is about $10^{-6}\ \rm{M_\odot/yr}$. The second dust-rich MB has recently been suggested as a Be/B[e]/LBV candidate. The gas lines of [\ion{Fe}{2}] as well as hot continuum components ($\sim300$ and $\sim1250$ K) arise from the inside of the MB while its outer shell emits a colder dust component ($\sim75$ K). The distance to the MB remains highly uncertain. Its mass is about $10^{-3}\ \rm{M_\odot}$ and its mass loss is about $10^{-5}\ \rm{M_\odot/yr}$.
\end{abstract}

\section{Introduction}

Evolved stars are significant contributors to the ionizing photons budget and kinetic energy released in the interstellar medium through their intense winds and radiation \citep{Crowther1998}. They are also the main sources of dust in the Galaxy and beyond. In the Milky Way and in the local universe, stars from the asymptotic giant branch (AGB) are the main producers of dust while red supergiants, novae and Wolf-Rayet stars (WR) contribute only traces of dust\citep[e.g.][]{Tielens2005,Matsuura2009}. Supernovae (SNe) have also been suggested as significant contributors to the dust production, at least at higher redshift, as large amount of dust are detected in young galaxies \citep[e.g.][]{Bertoldi2003,Dwek2009}. However, the dust yield from most sources, SNe in particular, is still controversial \citep[e.g.][and references therein]{Rho2009}. Evolved stars believed to be present in the Galaxy are still ``missing'' \citep{Shara1999}. Indeed, about 2500 planetary nebulae (PNe) are known in our Galaxy but only 224 central stars with spectral types are reported in the Strasbourg-ESO Catalogue of Galactic PNe \citep{Acker1992}. At least 1000 WRs stars are expected to be located within our Galaxy, but only 300 have been observed so far \citep{vanderHucht2001,vanderHucht2006}. Less than 50 luminous blue variables (LBVs) are known or identified as candidates \citep{Clark2005}. Therefore, it is essential to hunt for these evolved stars in order to determine their true nature and obtain an accurate census of the dust production in the Milky Way as well as in other galactic systems. 

\citet{Mizuno2010} discovered over 400 small ($<1\arcmin$) rings, bubbles, disks and shells from visual inspection of the MIPSGAL 24~$\mu$m mosaic images \citep{Carey2009}. These MIPSGAL bubbles (MBs) are pervasive through the entire Galactic plane in the mid-infrared. These objects span a large range of morphologies, sizes and fluxes. Their distribution is approximately uniform in Galactic latitude and longitude, and the average density is around 1.5 MBs per square degree. Only 15\% of the MBs exist in available catalogs from the VizieR and SIMBAD database. A large majority (8 out of 10) of the already known MBs are found in the MASH Catalogue of PNe \citep{Parker2006} and in the Catalogue of Galactic PNe \citep{Kohoutek2001}. Several supernova remnants (SNRs), WR stars, LBVs and emission line stars are also identified from published catalogues.

Extended emission in rings, shells and discs may be found (1) around main sequence and subgiant stars in the form of debris disc/rings associated with protoplanetary system, (2) around LBVs or associated with SNRs in the form of larger ring-type structure and (3) around PNe and symbiotic stars in the form of well deÞned ring structures \citep{Phillips2008}. It seems unlikely we can resolve structure around main sequence stars in the Galactic plane. The usual suspects for the MBs are thus mainly associated with stars in the late stages of their evolution. In these cases, the extended emission would arise from hot, small dust grains and/or from hot ionized gas in the stellar winds and ejecta. The MBs could account for the ``missing'' massive evolved stars. Because 85\% of these objects are yet unknown, there is a great potential for the discovery of several tens of new candidates. The identification of a statistical sample of these shells is thus of major importance for the study of stellar evolution, stellar mass-loss, dust life-cycle and ISM dynamics. Recently, \citet{Gvaramadze2010,Wachter2010} observed the central source detected in some MBs 24~$\mu$m images with optical and/or near-IR spectrometer and identify several new WR/LBV candidates. However, such observations can only be performed on MBs with a central source and does not answer the question on the origin of the MIPS 24~$\mu$m emission.

In this paper, we present mid-IR spectroscopic observations of 4 MBs obtained with the high resolution module ($R\sim600$) of the InfraRed Spectrograph \citep[IRS,][]{Houck2004} on board the Spitzer Space Telescope \citep{Werner2004} as part of our Spitzer Director's Discretionary Time (DDT) observational program. It is a unique set of observations\footnote{11 additional MBs have been observed in a different program with the low resolution module and will be presented in a subsequent paper.} to characterize the origin of the MIPS 24~$\mu$m emission of the MBs. Indeed, observations from the ground are limited in terms of sensitivity due to the extended nature of the MBs and the atmospheric transmission at those wavelengths, while from space, the next satellite with mid-IR capabilities (the James Webb Space Telescope) is not supposed to be launched before 2014. Meanwhile, the Stratospheric Observatory for Infrared Astronomy (SOFIA) is the best facility that can provide mid-IR capabilities, albeit with lower sensitivity than Spitzer. The structure of the paper is as follows. We present the observations and their reduction in section \ref{sec:obs} and \ref{sec:datared} respectively. We then analyze each MB separately, starting with the dust-poor MBs, in section \ref{sec:res}. We give our conclusions in section \ref{sec:ccl}.

\begin{table*}[!t]
\caption{List of the observed MBs. Nomenclature is from \citet{Mizuno2010}.}
\begin{center}
\begin{tabular}{l c r r r r r}
\hline
\hline
		& Name				& RA (J2000)	& DEC(J2000)	& $I_{24}$ (Jy)		& $r_{24}$ (\arcsec)	& $I_{70}$ (Jy) \\
\hline
MB4001	& MGE314.5619+00.1985	& 217.005		& -60.4739	& 0.36$\pm$0.01	& 15			& $< 0.2$ \\
MB4006	& MGE316.5508+00.5816	& 220.366		& -59.3494	& 0.98$\pm$0.02	& 21			& $< 0.3$ \\
MB3957	& MGE305.6514+00.3495	& 198.737		& -62.3980	& 0.59$\pm$0.02	& 19			& - \\
MB4121	& MGE337.5543+00.2200	& 249.178		& -46.9391	& 7.4$\pm$0.4	& 20			& 13 \\
\hline
\end{tabular}
\end{center}
\label{tab:coord}
\end{table*}

\begin{figure*}[!t]
\centering
\subfigure[]
	{\label{}
	\includegraphics[width=.24\linewidth]{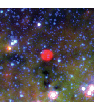}}
\subfigure[]
	{\label{}
	\includegraphics[width=.24\linewidth]{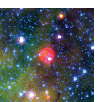}}
\subfigure[]
	{\label{fig:rgb_3957}
	\includegraphics[width=.24\linewidth]{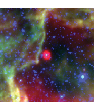}}
\subfigure[]
	{\label{fig:rgb_4121}
	\includegraphics[width=.24\linewidth]{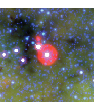}}
\caption{Three color image (blue is IRAC 4.5, green is IRAC 8.0 and red is MIPS 24~$\mu$m, 4.2\arcmin x4.2\arcmin) of (a) MB4001, (b) MB4006, (c) MB3957, (d) MB4121.}
\label{fig:rgb}
\end{figure*}

\begin{figure*}[!t]
\centering
\subfigure[]
	{\label{fig:m24_4001}
	\includegraphics[width=.24\linewidth]{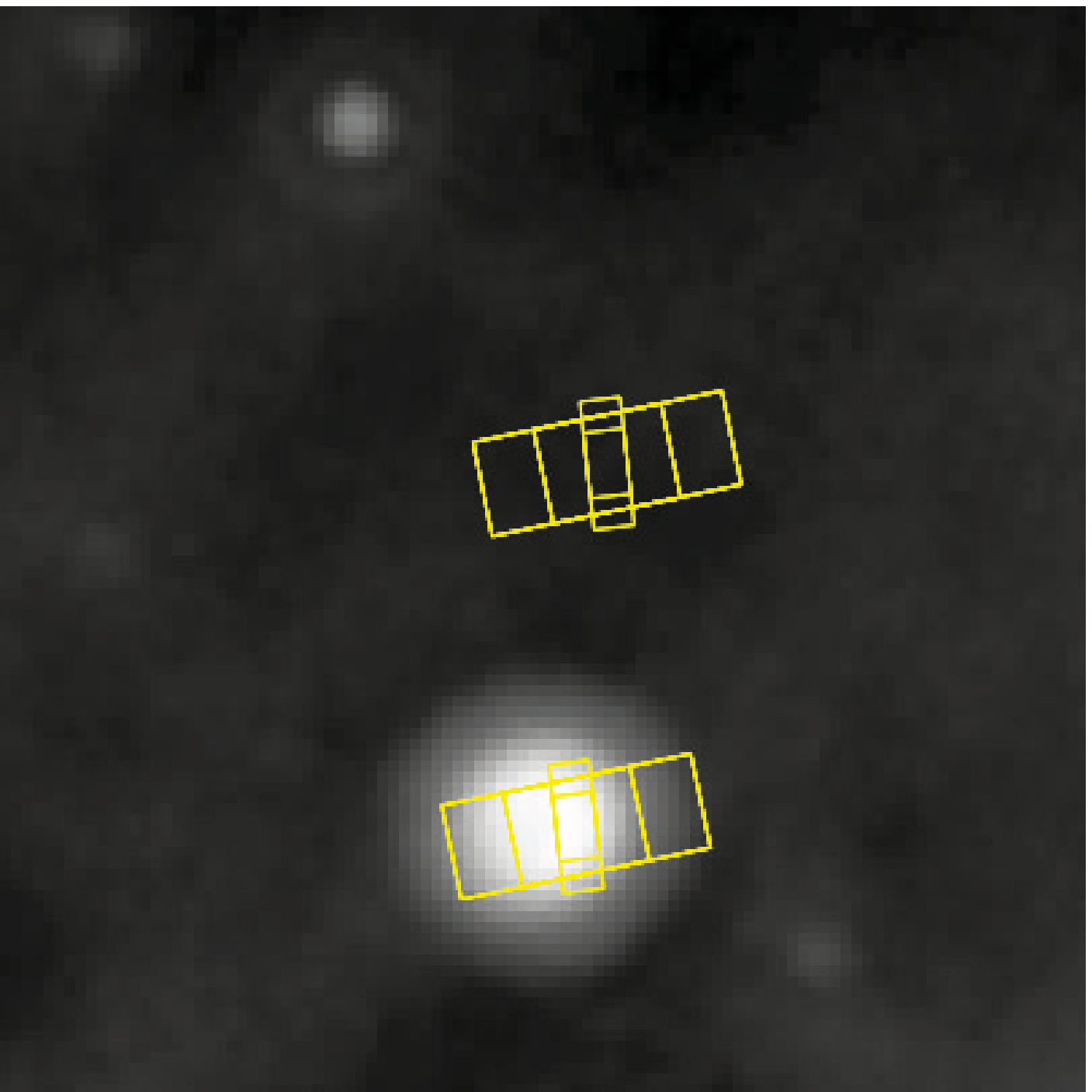}}
\subfigure[]
	{\label{fig:m24_4006}
	\includegraphics[width=.24\linewidth]{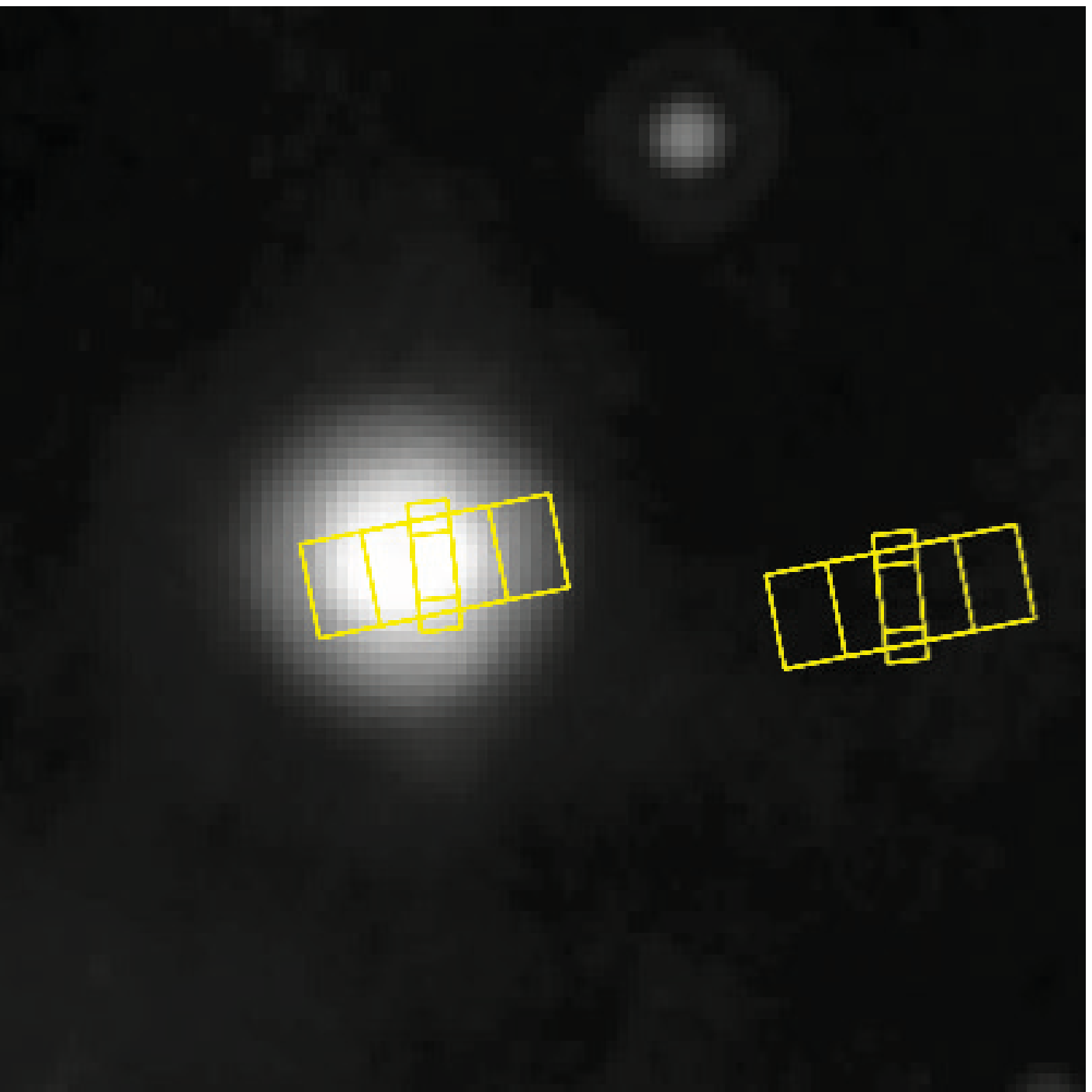}}
\subfigure[]
	{\label{fig:m24_3957}
	\includegraphics[width=.24\linewidth]{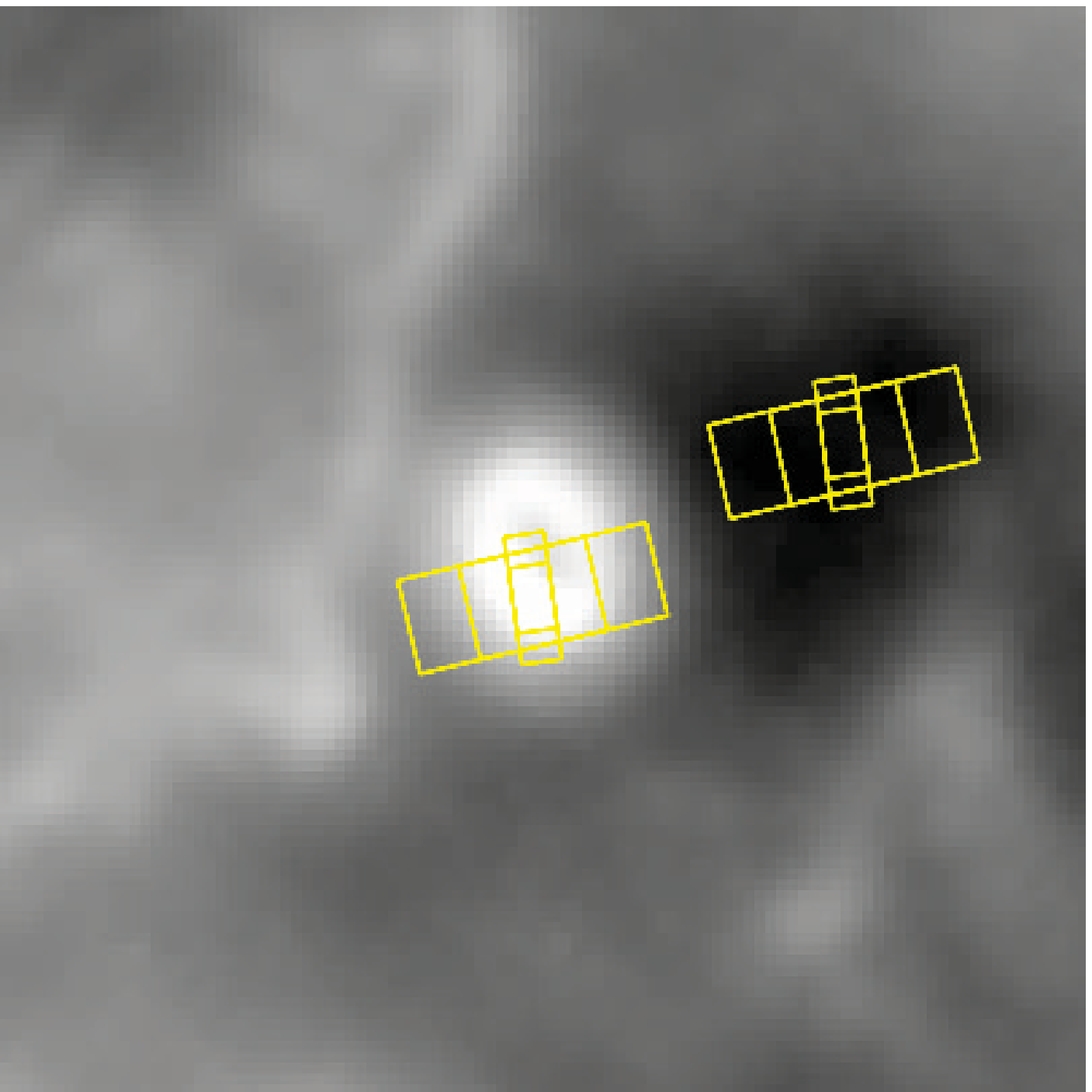}}
\subfigure[]
	{\label{fig:m24_4121}
	\includegraphics[width=.24\linewidth]{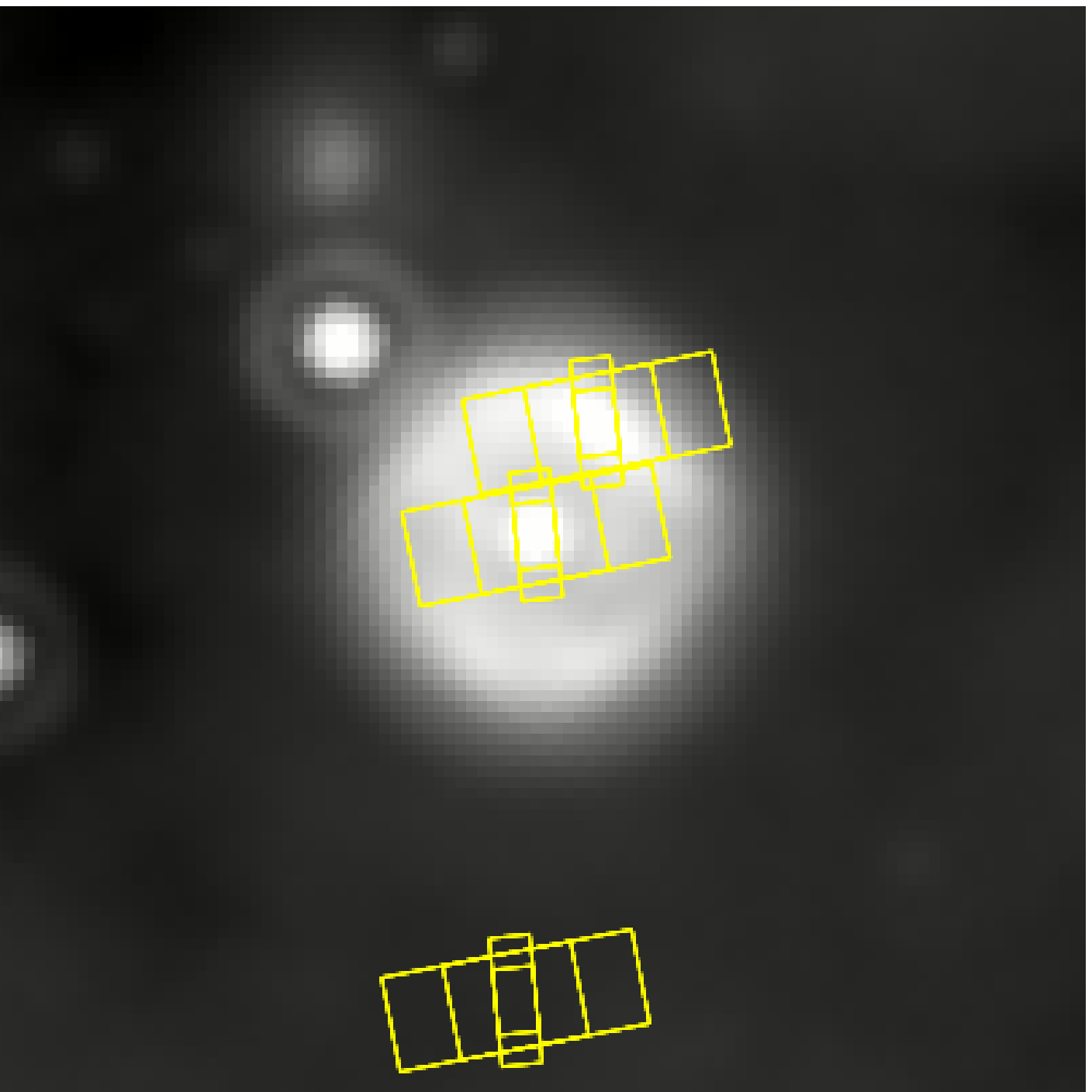}}
\caption{MIPS 24~$\mu$m image (2.1\arcmin x2.1\arcmin) with the positions of the IRS slit for (a) MB4001, (b) MB4006, (c) MB3957, (d) MB4121.}
\label{fig:aor}
\end{figure*}

\begin{figure*}[!t]
\centering
\subfigure[]
	{\label{fig:prof_4001}
	\includegraphics[angle=90,width=.24\linewidth]{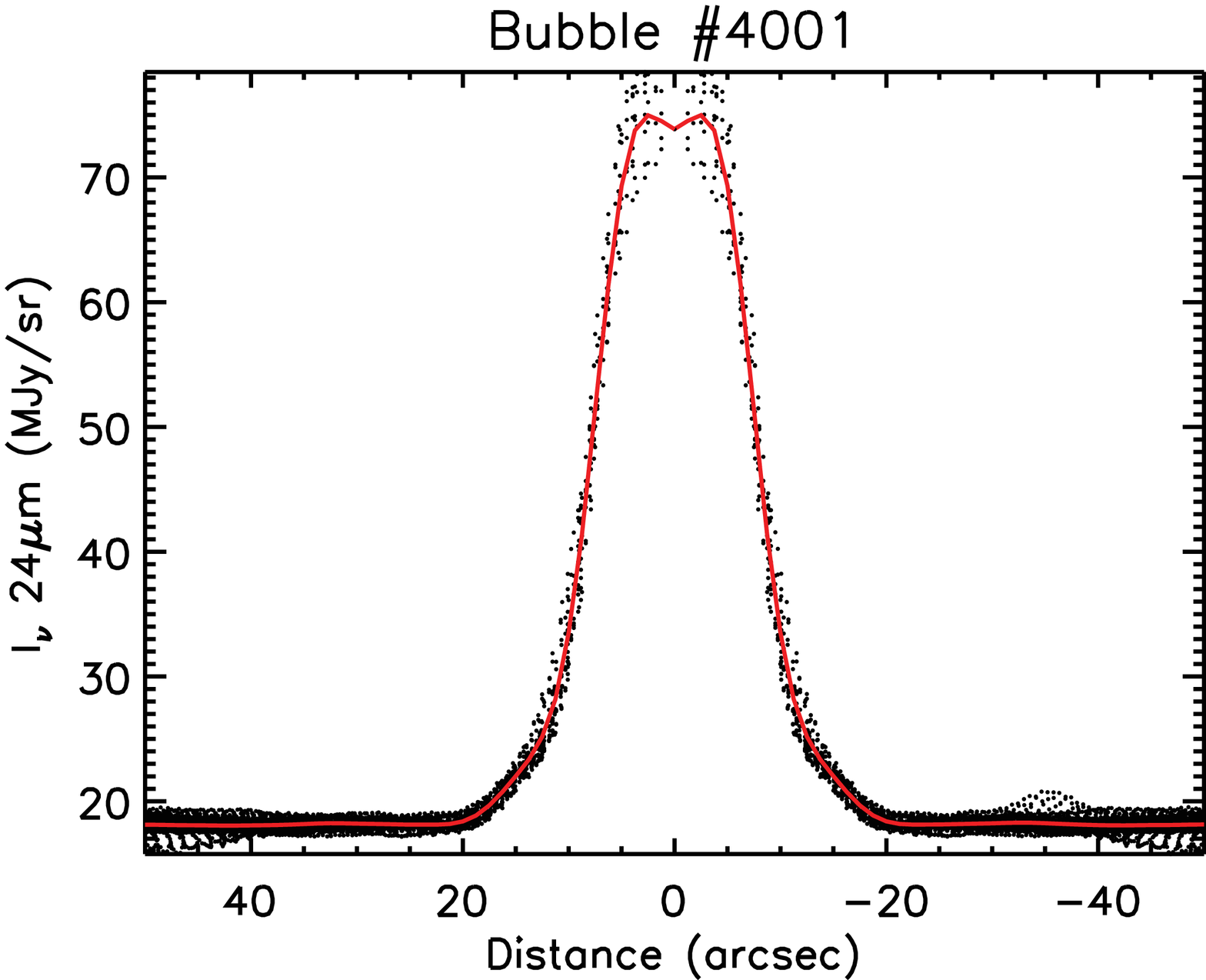}}
\subfigure[]
	{\label{fig:prof_4006}
	\includegraphics[angle=90,width=.24\linewidth]{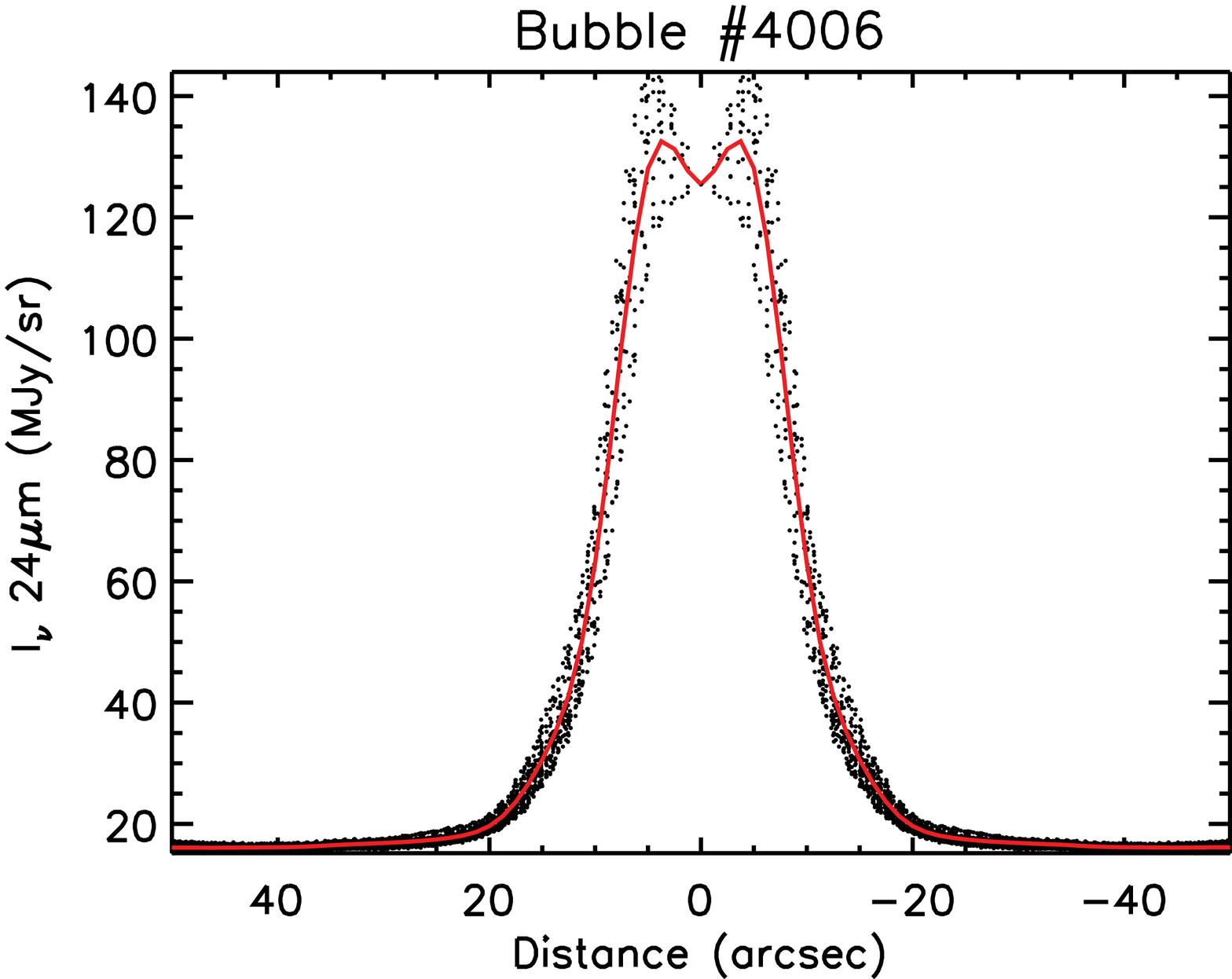}}
\subfigure[]
	{\label{fig:prof_3957}
	\includegraphics[angle=90,width=.24\linewidth]{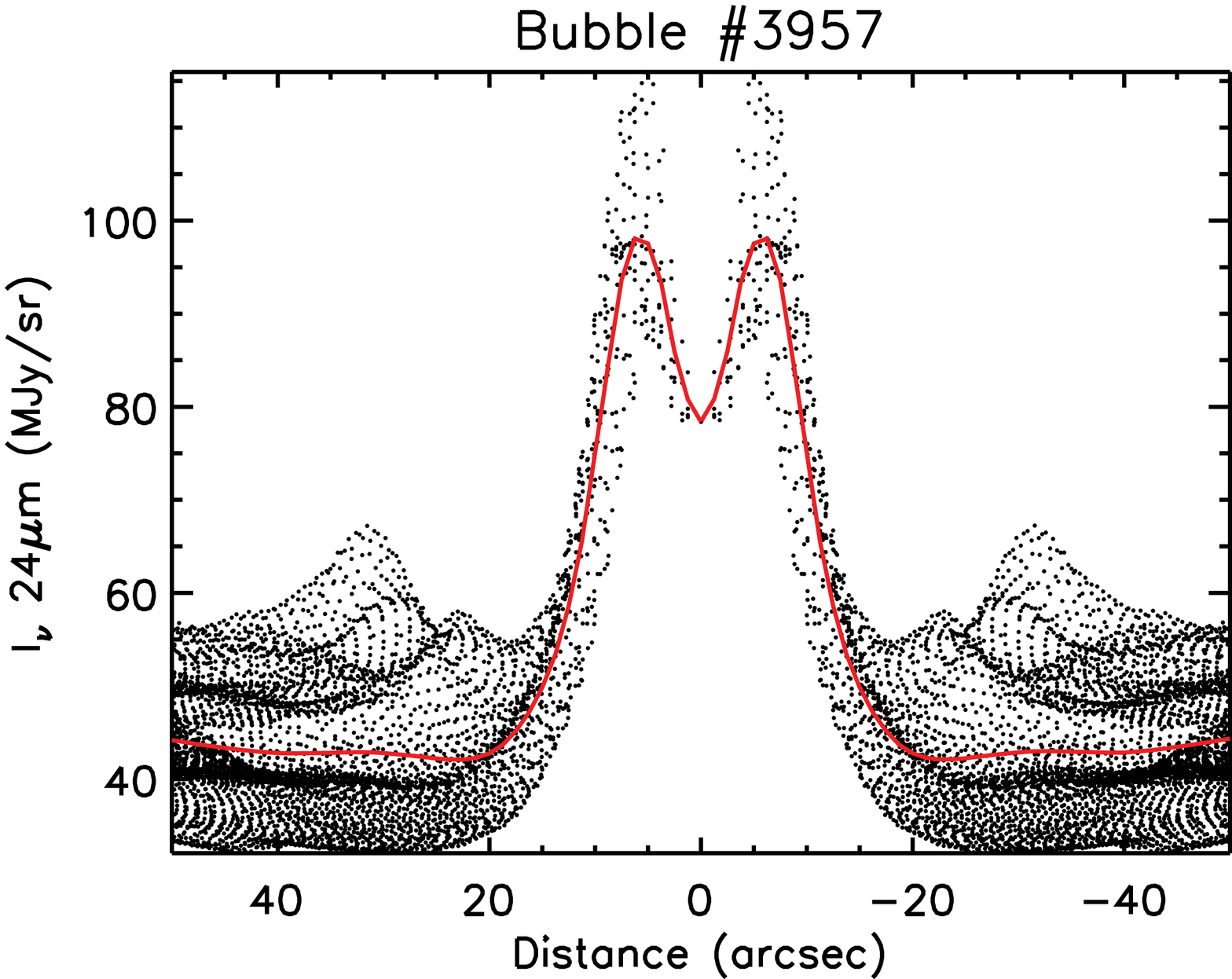}}
\subfigure[]
	{\label{fig:prof_4121}
	\includegraphics[angle=90,width=.24\linewidth]{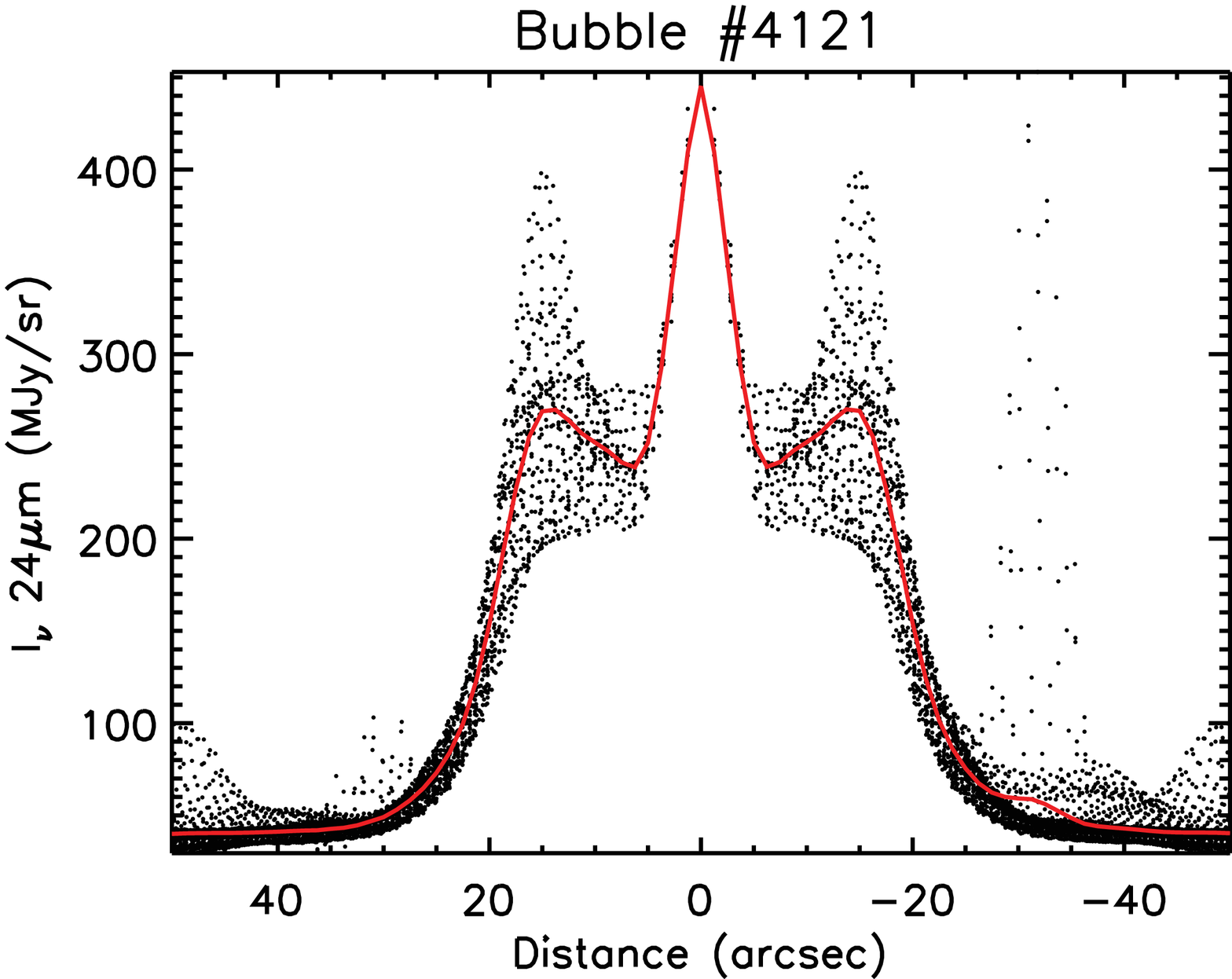}}
\caption{MIPS 24~$\mu$m radial profile of (a) MB4001, (b) MB4006, (c) MB3957, (d) MB4121. The black dots are all the pixels within 50\arcsec of the MB and the red line is the median as a function of the distance.}
\label{fig:prof}
\end{figure*}

\section{Observations}
\label{sec:obs}

The four MBs we present here have been observed on the 11th of April 2009, about a month before the end of Spitzer cryogenic mission. The four MBs are selected among the unidentified objects of the catalog that are visible at this time of the year by Spitzer. To make the best use of the one hour of observations we were granted, we chose four MBs that exhibit a significant MIPS 24~$\mu$m surface brightness and that sample different shell morphologies \citep[disk, shell and torus, see][for more details about the morphologies]{Mizuno2010}. However, we do not consider the 4 MBs presented in this paper represent to any extent a sample from which we could derive conclusions for the whole catalog of 428 MBs detected in the Galactic plane. The selected MBs are given in Table \ref{tab:coord} along with their identification number, name, coordinates, size and fluxes. Only MB4121 is detected in the MIPS 70~$\mu$m images and has a central source detected in the MIPS 24~$\mu$m images. Both MB3957 and MB4121 show a central source in 2MASS and IRAC images.

We select the short and long wavelengths high resolution ($R\sim600$) modules (SH from 9.9 to 19.6~$\mu$m, LH from 18.7 to 37.2~$\mu$m) in staring mode. For each MB, we have an ON position centered on the peak of the MIPS 24~$\mu$m extended emission and a dedicated OFF position for background subtraction that is observed in the same sequence as the ON position. For MB4121, we have an additional ON position centered on the central source detected in the MIPS 24~$\mu$m image and observed in the same sequence too. For the three brightest MB, we integrate for 4 cycles of 6 seconds in both LH and SH. For the faintest, MB4001, we integrate for 2 cycles of 30 seconds and 4 cycles of 14 seconds in SH and LH respectively. The position of the SH and LH slits are shown on top of MIPS 24~$\mu$m images in Figure \ref{fig:m24_4001}, \ref{fig:m24_4006}, \ref{fig:m24_3957} and \ref{fig:m24_4121}.

\section{Data reduction}
\label{sec:datared}

The first step of our data reduction corresponds to the co-addition of the different cycles of observations (basic calibrated data or BCD). We use for each pixel the median of the stack of BCDs. We build a \textit{master 2D image} for both on target and background positions separately.
We then subtract from each \textit{master image} the corresponding \textit{master background}. This process enables us to clean some of the rogue pixels. It also removes all the emission features from the Galactic ISM. In particular, we detect several weak emission features including H$_2$ lines and PAH features with the exact same intensity towards the background and the MBs. Consequently, these features which are commonly observed in a crowded and dense region such as the Galactic plane disappear from the MBs spectra after the background subtraction. Other gas lines are detected in the \textit{master background} with fluxes at least a 100 times fainter than in the \textit{master image}. We then perform another cleaning of the 2D images thanks to the IDL package developed by Jim Ingalls (SSC): IRSCLEAN\footnote{http://ssc.spitzer.caltech.edu/dataanalysistools/tools/irsclean/}.
We use SPICE\footnote{http://ssc.spitzer.caltech.edu/dataanalysistools/tools/spice/} to extract each background-subtracted spectrum. The IRS orders are overlapping each other over a few tenth of microns. We use these overlaps to manually trim the few first and/or last pixels of the order that are usually too noisy, unless both orders appear to be satisfying, in which case we average them.

The slit size of the 2 high resolution modules are different (11x22\arcsec\ for LH, 4x11\arcsec\ for SH) but both are similar to the characteristic size of our targets at 24~$\mu$m. As a consequence, we can assume our sources to behave neither as point source nor as extended source for the spectrum extraction purpose in SPICE. Therefore the most adapted calibration, which is wavelength dependent, is uncertain. We take into account the difference between those two extractions when estimating uncertainties on gas line fluxes. There is also a resulting mismatch (about a factor 2) between the SH and LH module spectra, even after correcting for the difference in the slit surface. The discrepancy might be caused by the fact the SH and LH slit do not probe exactly the same regions (see Fig.~\ref{fig:aor}). In order to correct these issues, we first scale the SH module to the LH module so that the dust continuum appears continuous in the overlapping wavelength range (19.0 to 19.5~$\mu$m). We scale the entire SH module, i.e. its continuum and the gas lines, the same way. We do this only for the two MBs that show a significant dust continuum in their spectrum. For the two other MBs, that exhibit a weak continuum, the match between the SH and LH modules might not be correct. While ratios between gas lines from the SH and the LH module might be inconsistent with models, they could be used to constrain the scaling factor between the two modules. We discuss this in section \ref{lab:dustfree}. We then scale the whole IRS spectrum of all four MBs so it matches the background subtracted MIPS 24~$\mu$m flux as seen through the LH slit and as measured in the MIPSGAL images. All the scaling factors that we apply are wavelength independent and are given in Table \ref{tab:corrfac}. The variations due to the nodding remain a negligible effect (about 1\%) except for the outer shell of MB4121. The scaling to MIPS 24~$\mu$m is closer to unity (1.26 to 1.35) with a point source extraction for the two MBs that are the most compact and closer to unity (0.87 to 0.91) with an extended source extraction for the two MBs that are the most extended.



\begin{table*}[ht]
\caption{Correction factors used to match the SH and LH modules in the case of the 2 MBs that show a continuum, and to match the MIPS 24~$\mu$m flux for all MBs. The correction factors are defined as: $I_\nu = CF_1 \times (I_\nu(SH) / CF_2 + I_\nu(LH))$ where $I_\nu$ is the final corrected spectrum, and $I_\nu(SH)$ and $I_\nu(LH)$ the SH and LH spectra extracted in SPICE, either with a point source (ps) or extended source (es) calibration. The given uncertainties result from the comparison between nodding positions.}
\begin{center}
\begin{tabular}{l r r r r}
\hline
\hline
		& $CF_1 (ps)$ & $CF_1 (es)$ & $CF_2 (ps)$ & $CF_2 (es)$ \\
\hline
MB4001	& $1.35\pm0.04$ & $1.91\pm0.05$ & - & -\\
MB4006	& $1.26\pm0.01$ & $1.77\pm0.01$ & - & -\\
MB3957	& $0.63\pm0.02$ & $0.87\pm0.02$ & $2.06\pm0.04$ & $1.44\pm0.03$ \\
MB4121 (central source)	& $0.65\pm0.01$ & $0.91\pm0.01$ & $2.6\pm0.1$ & $1.82\pm0.06$ \\
MB4121 (shell)	& $0.63\pm0.01$ & $0.87\pm0.01$ & $2.1\pm0.4$ & $1.5\pm0.3$ \\
\hline
\end{tabular}
\end{center}
\label{tab:corrfac}
\end{table*}

\section{Results}
\label{sec:res}

The IRS high resolution spectra of the four MBs in our sample are shown in figure \ref{fig:spec4001}, \ref{fig:spec4006}, \ref{fig:spec3957} and \ref{fig:4121_irs}. We classify them in two categories. The first corresponds to dust-poor bubbles and is comprised of MB4001 and MB4006. The second corresponds to dust-rich bubbles and is comprised of MB3957 and MB4121. The goal of this paper is to identify the nature of these objects thanks to the determination of some physical characteristics. To do so, we use MAPPINGS~III\footnote{Modelling And Prediction in PhotoIonised Nebulae and Gasdynamical Shocks, developed principally by Ralph Sutherland, Mike Dopita and Luc Binette and available at http://www.mso.anu.edu.au/~ralph/map.html}, a code of photoionization, and a simple model for the MBs.

\subsection{The use of MAPPINGS~III}
\label{lab:model}

In order to quantify the excitation conditions within our MBs, we use the MAPPINGS~III code of photoionization and model the MBs as isochoric spheres of gas heated by a central blackbody source and ionization bounded. We use a template provided with the MAPPINGS~III code. Details about the template are given in Table \ref{tab:mappings}. The MBs are modeled as isochoric sphere of gas illuminated by a source of $10^5$ km radius. We use otherwise solar abundances for the gas composition, except for iron. While iron may not be present in evolved low mass stars, it may play an important role in the cooling of the more massive ones. For similar reason, we also decide to run the MAPPINGS~III code with and without dust. Therefore we only add iron and dust in the model for the two MBs that show both iron lines and dust continuum in their IRS spectrum (see section \ref{lab:mb3957} and \ref{lab:mb4121}). The parameters we use for the dust are also given in Table \ref{tab:mappings}.

In the following sections, we compare the gas line fluxes predicted by the code to those observed in the MBs to constrain, among the many parameters available in the code, the electronic density of the bubble and the temperature of the inner source. To do so, we run MAPPINGS~III with density ranging from 100 to 15000 $\rm{cm^{-3}}$ and inner source temperature from $10^4$ to $3\times10^5$ K. The characterisation of the inner source then provides constraint on the nature of the MB. We also discuss how the presence of dust may affect the results.

\begin{table}[b]
\caption{Main parameters used for the MAPPINGS~III model}
\begin{center}
\begin{tabular}{l l}
\hline
\hline
Main parameters \\
\hline
Disable reactions	& No \\
Cosmic ray heating	& No  \\
Geometry			& Spherical  \\
Radius of Source	& 1.00D10 (cm) \\
Temperature of Source & $10^4 - 3\times10^5$ K \\
Structure			& Isochoric \\
Density                    & $100 - 1.5\times10^4\ \rm{cm^{-3}}$ \\
Filling Factor		& 1.0 \\
Integrate over whole sphere	& Yes  \\
Ionization balance	& Equilibrium \\
Photon absorption fraction	& 0.1 \\
Ionization bounded	& Applies to H \\
				& up to 5\% ionization \\
\hline
\hline
Dust parameters \\
\hline
Shattering grain profile \\
Power law & -3.3 \\
Graphite grain radius & 40-1600 \\
Silicate grain radius & 40-1600 \\
Graphite grain density  & 1.8 \\
Silicate grain density & 3.5 \\
Include PAH molecules & Yes \\
Fraction of Carbon Dust &  \\
\multicolumn{1}{r}{Depletion in PAHs} & 0.05 \\
\multicolumn{2}{l}{Q-value QHDH $<$ Value PAH photodissociation parameter} \\
PAH switch on Value & 1000 \\
Destroy all C grains with PAHs & Yes \\
\hline
\end{tabular}
\end{center}
\label{tab:mappings}
\end{table}%

\subsection{Dust-poor objects: MB4001 and MB4006}
\label{lab:dustfree}

\begin{figure*}[!t]
\centering
\subfigure[]
	{\label{fig:spec4001}
	\includegraphics[angle=90,width=.45\linewidth]{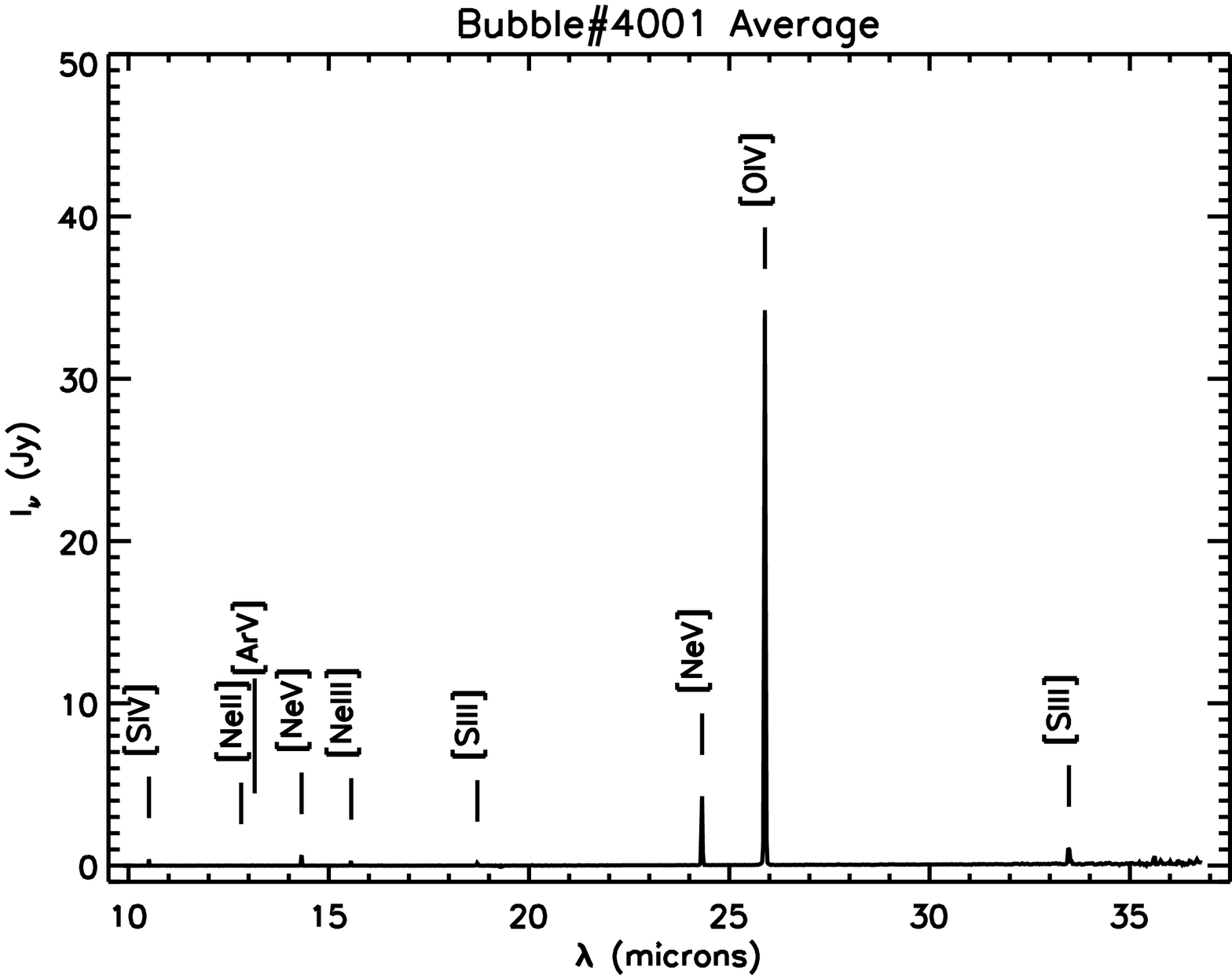}}
\subfigure[]
	{\label{fig:spec4006}
	\includegraphics[angle=90,width=.45\linewidth]{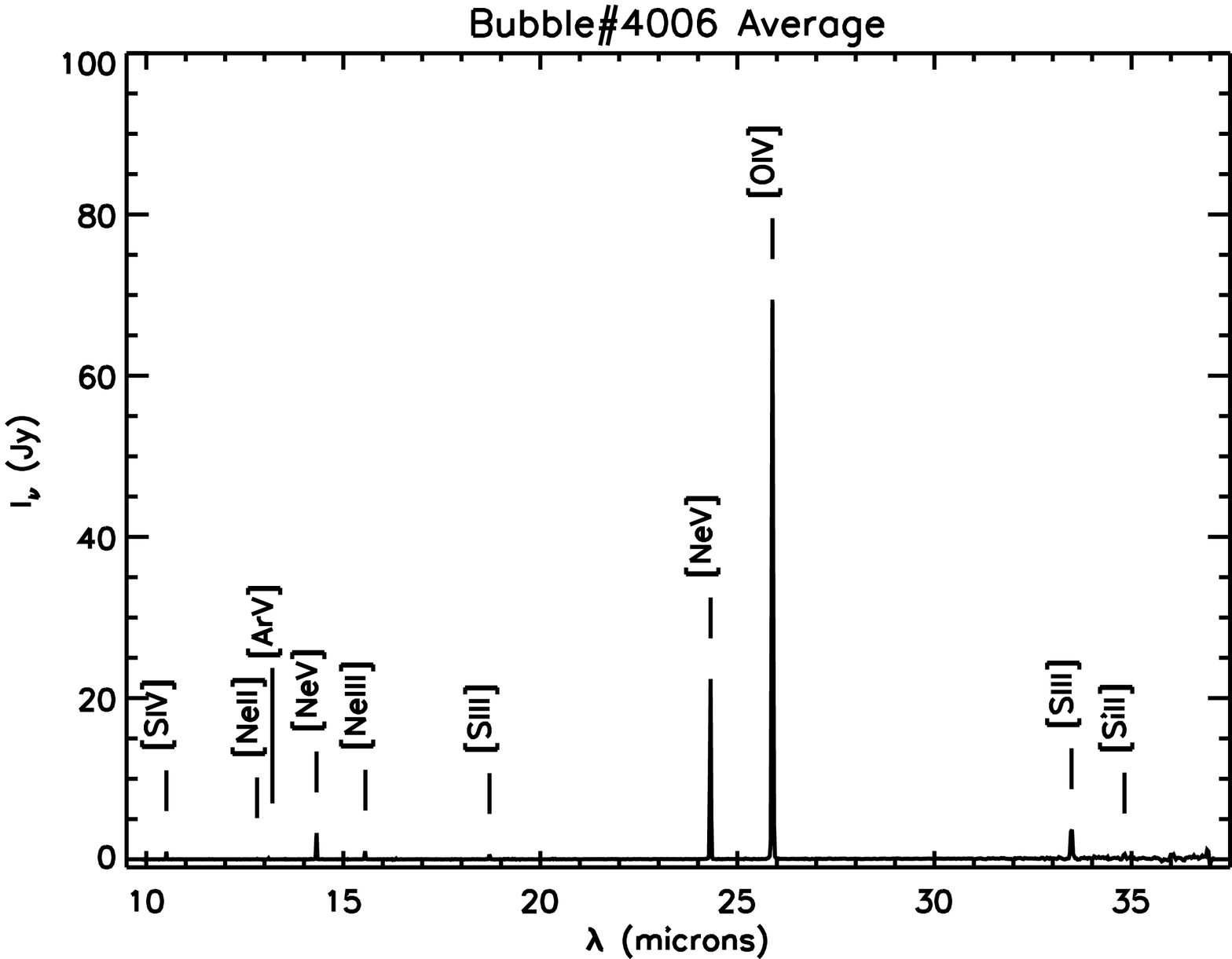}}
\caption{IRS spectra of (a) MB4001 and (b) MB4006 with detected gas lines indicated.}
\label{}
\end{figure*}

The first two MBs of our program are morphologically very similar at 24~$\mu$m (see Figure \ref{fig:m24_4001} and \ref{fig:m24_4006}). Their radial profile at 24~$\mu$m exhibit only a slight difference at the center: MB4001 is flatter and MB4006 has a deeper central hole. For this reason they have been classified as a ``disk'' and a ``ring'' respectively in \citet{Mizuno2010}. Their IR spectrum are also very similar and are shown in Figure \ref{fig:spec4001} and \ref{fig:spec4006}. They have a very weak continuum and many high ionization gas lines ([\ion{O}{4}] 25.9~$\mu$m, [\ion{Ne}{5}] 24.3 and 14.3~$\mu$m, [\ion{Ne}{3}] 15.5~$\mu$m, [\ion{Ar}{5}] 13.1~$\mu$m, [\ion{S}{3}] 18.7 and 33.5~$\mu$m). The flux of these lines are given in Table \ref{tab:flux4001} and \ref{tab:flux4006}. We first identify those two MBs as planetary nebulae candidates and then constrain some of their properties.

\subsubsection{Identification as PNe candidates}
\label{lab:dustfree_ident}

These two MBs share common properties with SMP83 in the LMC \citep{BernardSalas2004} and M1-42 in the Galactic center \citep{Pottasch2007}, two known planetary nebulae (PNe). They are also very similar to the object discovered in Cepheus by \citet{Morris2006} and suggested as a dust-free supernova remnant: detection within MIPS 24~$\mu$m bandwidth only (neither in IRAC 3.6 to 8.0~$\mu$m nor in MIPS 70 and 160~$\mu$m), high excitation gas lines ([\ion{O}{4}] 25.9~$\mu$m, [\ion{Ne}{5}] 24.3 and 14.3~$\mu$m), very little dust continuum. However, the gas lines ratios are significantly different. In particular, our dust-poor MBs exhibit I([\ion{Ne}{5}] 24.3~$\mu$m) $\gg$ I([\ion{Ne}{3}] 15.5~$\mu$m) and I([\ion{Ne}{5}] 14.3~$\mu$m) $\gg$ I([\ion{Ne}{3}] 15.5~$\mu$m), contrary to the Cepheus bubble.

To account for the lack of dust emission, \citet{Morris2006} suggest that very little interaction between the shell and the ISM has occurred and that the shell has a very high gas-to-dust ratio. They rule out AGB, post-AGB and PN shell based on the following arguments: (1) the electronic density $N_e$ they derive from [\ion{S}{3}] 33.5~$\mu$m to [\ion{S}{3}] 18.7~$\mu$m does not depend on the radius while multilayered shell would create large variations in $N_e$, (2) typical IR spectra of PNe show PAHs features and higher ionized gas species, (3) the nebular gas spectrum should contain hydrogen lines tracing the hydrogen lifted away from the surface of the star in its AGB or post-AGB phases prior to the star emerging the H-deficient white dwarf cooling track. However, \citet{Fesen2010} obtained narrow and broad passband optical images as well as low dispersion optical spectra of the Cepheus bubble. They reject the dust-free SNR interpretation based on the absence of high-velocity features, the presence of relatively strong [\ion{N}{2}] emission and the lack of detected [\ion{S}{2}] emission, which would indicate the presence of shock-heated gas. Additionally, no coincident X-ray or nonthermal radio emission has been detected. They suggest it is more likely a faint high-excitation PN, lying at a distance $\sim$ 2.5 kpc with strong [\ion{O}{4}] emission rather than dust continuum emission dominating in the MIPS 24~$\mu$m images. They have identified a possible central star candidate with $m_{r'} \simeq 22.4$.

Furthermore, while the presence of PAHs features in PNe has been reported by many authors: in NGC7027 \citep{Beintema1996}, Magellanic Clouds PNe \citep[several examples in ][]{Stanghellini2007, BernardSalas2008, BernardSalas2009} and Galactic Bulge PNe \citep[one example, PNG351.2+05.2 in ][]{Gutenkunst2008}, there are also examples of PAH free PNe, sometimes in the same references: SMP83 in the Large Magellanic Cloud (LMC) \citep{BernardSalas2004}, a rather faint PN in the direction of the Galactic center \citep[M1-42 ][]{Pottasch2007}, Magellanic Clouds PNe \citep[several examples in][]{Stanghellini2007, BernardSalas2008, BernardSalas2009} and Galactic Bulge PNe \citep[all but one in][]{Gutenkunst2008}. Among those PNe that show PAH-free IR spectrum, some present a significant continuum at wavelengths longer than 15~$\mu$m or crystalline silicate features \citep{Stanghellini2007, Gutenkunst2008} while others hardly show any continuum or broad features from 10 to 35~$\mu$m. MB4001 and MB4006 as well as the bubble of \citet{Morris2006} belong to this last class of object, which may be called dust-poor PNe. Other examples are SMP13, SMP35, SMP40 and SMP83 in the LMC \citep{BernardSalas2004, BernardSalas2008} as well as M1-42 toward the Galactic center \citep{Pottasch2007}.

Finally, \citet{Groves2008} studied the neon and sulfur emission lines for a wide variety of objects (e.g. galaxies, PNe, HII regions) from the Spitzer and ISO archives. Their diagram of [\ion{Ne}{3}]~15.6~$\mu$m /[\ion{Ne}{2}]~12.8~$\mu$m vs [\ion{S}{4}]~10.5~$\mu$m /[\ion{S}{3}]~18.7~$\mu$m shows a tight correlation. The two dust-poor MBs would fall on the high end of this correlation, where the PNe, which represent the most isolated population in their sample, are the most numerous.

We conclude that MB4001 and MB4006 are undoubtedly PNe candidates.

\begin{table}[b]
\caption[ ]{\label{tab:flux4001} Gas lines detected in the IRS spectrum of MB4001. Units are $10^{-20}\ \rm{W.cm^{-2}}$. The lines in the SH module may be underestimated by a factor of a few (see text in section \ref{lab:dustfree_charac} for details).}
\begin{center}
\begin{tabular}{l r@{.}l r@{.}l}
\hline
\hline
Line & \multicolumn{2}{c}{Flux} & \multicolumn{2}{c}{Uncertainty} \\
\hline
~[\ion{S}{4}] 10.51~$\mu$m &       2&09 &      0&02 \\
~[\ion{Ne}{2}] 12.81~$\mu$m &      0&015 &      0&001 \\
~[\ion{Ar}{5}] 13.10~$\mu$m &      0&092 &      0&004 \\
~[\ion{Ne}{5}] 14.32~$\mu$m &      2&7 &      0&2 \\
~[\ion{Ne}{3}] 15.56~$\mu$m &      0&96 &      0&09 \\
~[\ion{S}{3}] 18.71~$\mu$m &      0&49 &      0&07 \\
\hline
~[\ion{Ne}{5}] 24.32~$\mu$m &      6&55 &      0&07 \\
~[\ion{O}{4}] 25.89~$\mu$m &     61&7 &      0&2 \\
~[\ion{S}{3}] 33.48~$\mu$m &      1&33 &      0&09 \\
\hline
\end{tabular}
\end{center}
\end{table}

\begin{table}[b]
\caption[ ]{\label{tab:flux4006} Gas lines detected in the IRS spectrum of MB4006. Units are $10^{-20}\ \rm{W.cm^{-2}}$. The lines in the SH module may be underestimated by a factor of a few (see text in section \ref{lab:dustfree_charac} for details).}
\begin{center}
\begin{tabular}{l r@{.}l r@{.}l}
\hline
\hline
Line & \multicolumn{2}{c}{Flux} & \multicolumn{2}{c}{Uncertainty} \\
\hline
~[\ion{S}{4}] 10.51~$\mu$m	& 5&05	& 0&05 \\
~[\ion{Ne}{2}] 12.81~$\mu$m	& 0&19	& 0&01 \\
~[\ion{Ar}{5}] 13.10~$\mu$m	& 0&46	& 0&02 \\
~[\ion{Ne}{5}] 14.32~$\mu$m	& 13&1	& 0&9 \\
~[\ion{Ne}{3}] 15.56~$\mu$m	& 3&3	& 0&3 \\
~[\ion{S}{3}] 18.71~$\mu$m	& 1&5	& 0&2 \\
\hline
~[\ion{Ne}{5}] 24.32~$\mu$m	& 34&6	& 0&3 \\
~[\ion{O}{4}] 25.89~$\mu$m	& 123&0	& 0&5 \\
~[\ion{S}{3}] 33.48~$\mu$m	& 4&9	& 0&4 \\
~[\ion{Si}{2}] 34.82~$\mu$m	& 0&83	& 0&07 \\
\hline
\end{tabular}
\end{center}
\end{table}

\subsubsection{Characterization}
\label{lab:dustfree_charac}

\begin{figure}[t]
	\centering
	\includegraphics[angle=90,width=\linewidth]{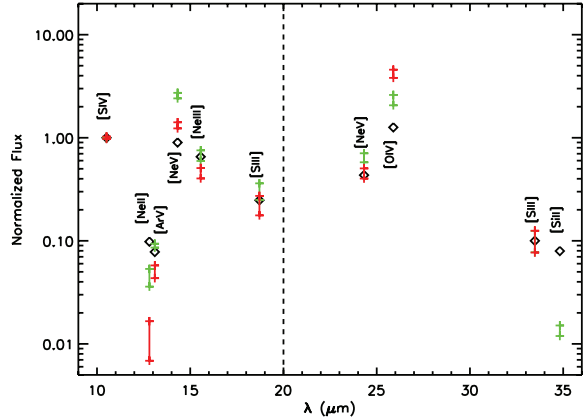}
	\caption{Comparison of gas line fluxes between MB4001 (red), MB4006 (green) and SMP83 from \citet{BernardSalas2004} in diamonds. Fluxes are normalized to [\ion{S}{4}]~10.5~$\mu$m $= 1$ in SH module (left of the dashed line) and to [\ion{S}{3}] 33.5~$\mu$m $= 0.1$  in LH module (right of the dashed line).}
	\label{fig:PNcomp}
\end{figure}

Even though the examples of dust-poor PNe in the literature are not numerous, MB4001 and MB4006 both show strong similarities with SMP83 \citep{BernardSalas2004} and, to a lesser extend, M1-42 \citep{Pottasch2007} and Sh 2-188 \citep{Chu2009}. We compare gas lines fluxes of the two MBs, normalized to [\ion{S}{4}]~10.5~$\mu$m in the SH module and to [\ion{S}{3}]33.5~$\mu$m in LH module, to those of SMP83 in Figure \ref{fig:PNcomp}. We normalize and discuss both modules separately, as some mismatch between the modules might remain (see section \ref{sec:datared}). Within the SH module, we observe that the [\ion{S}{4}]~10.5~$\mu$m to [\ion{S}{3}]~18.7~$\mu$m line ratio appears to be very similar within the three objects. The same can be told about the [\ion{Ar}{5}]~13.1~$\mu$m line relative to the [\ion{S}{4}]~10.5~$\mu$m line. The other lines within the SH modules are three Neon lines at three different ionization levels ([\ion{Ne}{2}]~12.8~$\mu$m, [\ion{Ne}{3}]~15.5~$\mu$m and [\ion{Ne}{5}]~14.3~$\mu$m). Based on their relative fluxes MB4001 and MB4006 appear to be more highly ionized than SMP83. Within the LH module, all the gas lines arise from different elements. On the one hand, the ratio between [\ion{Ne}{5}]~24.3~$\mu$m and [\ion{S}{3}]33.5~$\mu$m is consistent within the three objects. On the other hand, the [\ion{Si}{2}]34.8~$\mu$m line is much weaker, relative to [\ion{Ne}{5}]~24.3~$\mu$m, in MB4006 than in SMP83. It is not detected in MB4001. Additionally, the [\ion{O}{4}]~25.9~$\mu$m appears to be stronger in the MBs than in SMP83. We note that it's again within MB4001 that excitation, as traced by the [\ion{O}{4}]~25.9~$\mu$m line, seems to be the strongest.

\begin{figure}[t]
	\centering
	\includegraphics[angle=90,width=\linewidth]{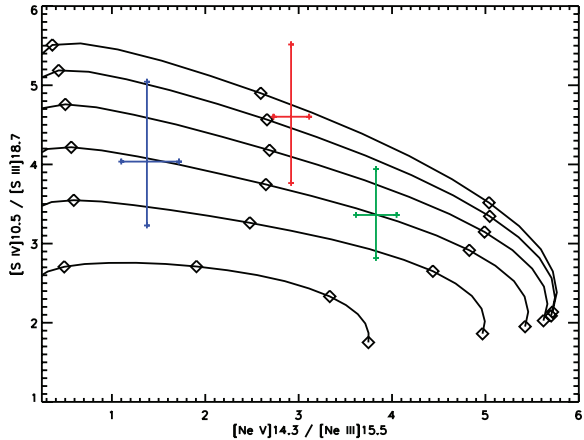}
	\caption{[\ion{S}{4}]~10.5~$\mu$m to [\ion{S}{3}]~18.7~$\mu$m ratio as a function of [\ion{Ne}{5}]~14.3~$\mu$m to [\ion{Ne}{3}]~15.5~$\mu$m ratio from the MAPPINGSIII model without dust for different set of temperature of the inner source and density of the isochoric sphere of gas. From bottom to top, the curves are representing densities of 1000, 3000, 5000, 7000, 9000 and 10000 cm$^{-3}$. From left to right, diamonds are for temperatures of 150000, 200000, 250000 and 300000 K. MB4001 and MB4006 (this paper) and SMP83 \citep{BernardSalas2004} are indicated with the red, green and blue crosses respectively.}
	\label{fig:PNdiag}
\end{figure}

In order to quantify the excitation conditions within the two dust-poor MBs, we use MAPPINGS~III with the parameters presented in Table \ref{tab:mappings} and without dust. We use the results from these models to compare the IR gas line fluxes to those observed in the dust-poor MBs and then constrain the temperature of the inner source of radiation and the gas density of the MB. In order to avoid the effect of any remaining LH/SH mismatch and any dependence on the chemical composition of the gas, we use as diagnostics the [\ion{S}{4}]~10.5~$\mu$m to [\ion{S}{3}]~18.7~$\mu$m and [\ion{Ne}{5}]~14.3~$\mu$m to [\ion{Ne}{3}]~15.5~$\mu$m line ratios. Figure \ref{fig:PNdiag} shows the evolution of both these ratios as predicted by our models for various temperature and density. Measurements for MB4001, MB4006 and the planetary nebula SMP83 are also shown. The line ratios of SMP83 are well reproduced by a density of $5000\pm2500\ \rm{cm^{-3}}$ and a source temperature of about $170000\pm10000\ \rm{K}$. The density we estimate is in agreement with previous measurements \citep[][and references therein]{BernardSalas2004}. The temperature of the central source is the same as that estimated by \citet{Dopita1993} with MAPPINGS 2 and a multizone model combining an optically thick ring with optically thin extensions. They found densities of 1000 and 3600 $\rm{cm^{-3}}$ in the thin and thick components respectively. The required densities are $10000\pm5000\ \rm{cm^{-3}}$ and $5000\pm1500\ \rm{cm^{-3}}$ in MB4001 and MB4006 respectively, which are similar to that of SMP83. The inner source temperature is significantly hotter though. The IR gas lines of MB4001 are in agreement with $T = 205000\pm5000 \ \rm{K}$ while those of MB4006 are in agreement with  $T = 225000\pm10000 \ \rm{K}$. Those temperature are on the high-end of what is found in published examples of PNe.

Using these parameters, we are able to predict ratios between SH and LH lines (e.g. [\ion{S}{3}]~33.5~$\mu$m to [\ion{S}{3}]~18.7~$\mu$m and [\ion{Ne}{5}]~14.3~$\mu$m to [\ion{Ne}{5}]~24.3~$\mu$m) and thus, by comparing them to the observed fluxes, estimate the scaling factor that should be applied to match the SH module with the LH module. For MB4001, the low end of the density range ($\sim5000\ \rm{cm^{-3}}$) leads to an SH scaling factor of about 4.5 and 7.0 using the [\ion{S}{3}] and [\ion{Ne}{5}] line ratios respectively, while the high end of the density range ($\sim15000\ \rm{cm^{-3}}$) leads to a scaling factor of about 8.4 and 17.0 using the [\ion{S}{3}] and [\ion{Ne}{5}] line ratios respectively. For MB4006, those values are 3.7 and 5.5 at low densities ($\sim3500\ \rm{cm^{-3}}$) and 6.3 and 15.0 at high densities ($\sim6500\ \rm{cm^{-3}}$). We decide not to apply any scaling factor to the SH module given such a large uncertainty. However, we draw the reader's attention to the underestimation, by a factor of a few, of the SH lines fluxes given in Table \ref{tab:flux4001} and \ref{tab:flux4006}.

From the comparison of the gas line fluxes with those of SMP83, we also give a rough estimate of the distances to MB4001 and MB4006. We use only LH gas lines to perform this comparison. The [\ion{Ne}{5}] 24.3 and [\ion{O}{4}] 25.89~$\mu$m are respectively 4.3 and 14 times brighter in MB4001 than in SMP83. Those values are 23 and 28 for MB4006. The distance to SMP83 is 50 kpc. To scale the emission lines of MB4001 and MB4006 adequately requires them to be between 18 and 200 times closer (0.25 and 2.8 kpc) and between 530 and 780 times closer (0.06 and 0.1 kpc) respectively, depending on the gas line used for the scaling. These estimations assume the three PNe have similar physical properties. However, since we know the two MBs are significantly hotter than SMP83, those distances might be lower-limits. Another estimate can be provided by the apparent sizes of the MBs. At a distance of 50 kpc, SMP83 is smaller than 1\arcsec, at least 15 times smaller than MB4001 and 21 times smaller than MB4006. Therefore, the two MBs would be at a distance of at most 3.3 and 2.4 kpc respectively. At those distances, the MBs would have a physical size of about 0.1 to 0.2 pc which is satisfying. Moreover, at an arbitrary distance of 2 kpc, a white dwarf of 200000 K surface temperature and 5000 km in radius has an apparent magnitude of at least 20 from $V$ band to IRAC and MIPS channels. Therefore such temperatures and distances are consistent with the non-detection of the inner source of the dust-poor MBs, especially in the Galactic plane where the extinction is high.

Finally, we discuss the very limited detection of dust emission in the MBs spectra. No PAH features are detected at short wavelengths in the IRS spectrum and the IRAC images. The MIPS 70~$\mu$m broadband images (not shown here) do not present any trace of colder dust. The MIPS 24~$\mu$m is mainly accounted for by the [\ion{Ne}{5}] 24.3~$\mu$m and [\ion{O}{4}] 25.9~$\mu$m lines. In the case of MB4001, a weak continuum ($<0.1$ Jy) is detected in the LH range of the spectrum but its contribution to the total MIPS 24~$\mu$m flux is only about 30\%. In the case of MB4006, the continuum is hardly detected and its contribution to the MIPS 24~$\mu$m emission is less than 15\%. The detection of a weak [\ion{Si}{2}] line in MB4006 indicates that a faint PDR might still be present in this object. As we already mentioned in section \ref{lab:dustfree_ident}, examples of such dust-poor PNe exist in the literarure. The most likely explanation is that the dust and the PDR have been destroyed by the hard radiation field of the inner source. It has been found that larger, more evolved, PNe usually exhibit less dust continuum and more [\ion{O}{4}] gas line emission in their MIR spectrum \citep{Stanghellini2007, Chu2009} as the amount of ionized gas increases at the expense of the molecular and atomic material \citep{Huggins1996, BernardSalas2005}. In this global picture, MB4001 and MB4006 would be template of old PNe.


\subsection{Dust-rich objects}

The other two objects in our sample share at least two properties: (1) a IR central source is detected in several Spitzer channels and (2) their mid-IR spectra is dominated by a continuum. However, these two objects are as much different from the first two as they are from each other, both in terms of IR morphology and spectroscopy. We thus present them separately.

\subsubsection{MB3957}
\label{lab:mb3957}

\begin{figure}[t]
	\centering
	\includegraphics[width=\linewidth]{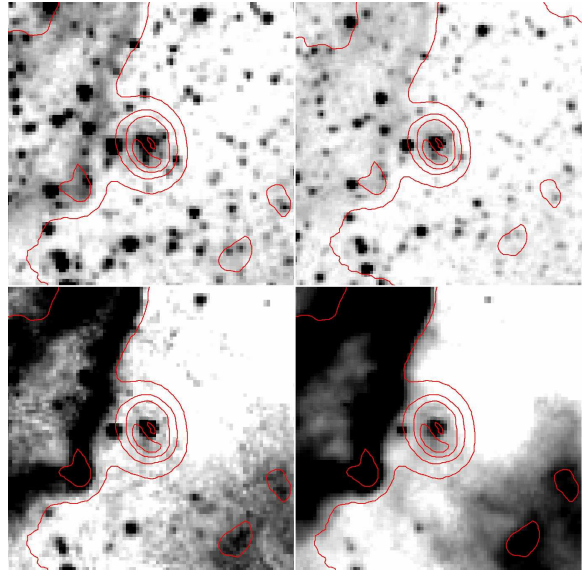}
	\caption{IRAC 3.6 (top-left), 4.5 (top-right), 5.8 (bottom-left) and 8.0~$\mu$m (bottom-right) images (2\arcmin x2\arcmin) with contours of MIPS 24~$\mu$m of MB3957.}
	\label{fig:3957_irac}
\end{figure}


\begin{figure}[t]
	\centering
	\includegraphics[angle=90,width=\linewidth]{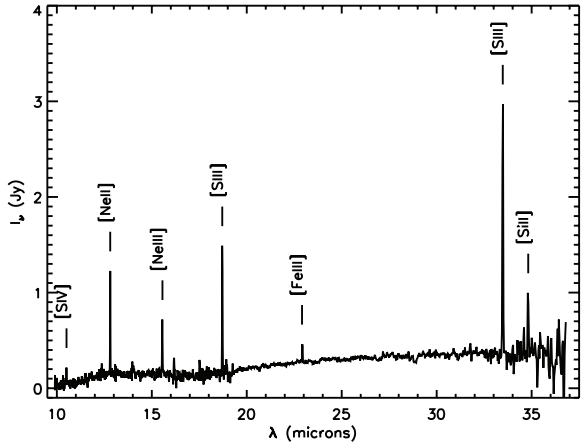}
	\caption{IRS spectrum of MB3957.}
	\label{fig:spec3957}
\end{figure}

\begin{table}[b]
\caption[ ]{\label{tab:flux3957} Line fluxes detected in the IRS spectrum of MB3957. Units are $10^{-20}\ \rm{W.cm^{-2}}$. The extinction corrected line fluxes results from the whole SED fitting (see section \ref{sec:mb3957_wholefit}).}
\begin{center}
\begin{tabular}{l r@{.}l@{$\pm$}r r@{.}l@{$\pm$}l}
\hline
\hline
Line & \multicolumn{3}{c}{Flux} & \multicolumn{3}{c}{Corrected flux} \\
\hline
~[\ion{S}{4}]~10.51$\mu$m &    0&9 &  0.8 &  4&4 &  3.9 \\
~[\ion{Ne}{2}]~12.81$\mu$m &  3&7 &  0.7 &  7&0 &  1.0 \\
~[\ion{Ne}{3}]~15.56$\mu$m &  2&3 &  0.5 &  5&0 &  1.0 \\
~[\ion{S}{3}]~18.71$\mu$m &  3&7 &  0.8 & 11&0 &  2.0 \\
~[\ion{Fe}{3}]~22.92$\mu$m &  0&4 &  0.3 &  1&0 &  0.9 \\
~[\ion{S}{3}]~33.48$\mu$m &  4&0 &  1.0 &  7&0 &  2.0 \\
~[\ion{Si}{2}]~34.82$\mu$m &  0&9 &  1.2 &  1&5 &  2.0 \\
\hline
\end{tabular}
\end{center}
\end{table}

The MIPS 24~$\mu$m morphology of MB3957 is that of a torus. A lack of MIR emission at the center is obvious from the MIPS 24~$\mu$m image and radial profile (see Fig.~\ref{fig:m24_3957} and \ref{fig:prof_3957}). Figure \ref{fig:3957_irac} shows the IRAC observations of MB3957, from the GLIMPSE survey \citep{Benjamin2003} with contours from the MIPS 24~$\mu$m observation. Extended emission associated with MB3957 is visible at every IRAC wavelengths, though the presence of nearby bright diffuse emission prevents us from an unequivocal conclusion. A point source is detected at the center of MB3957 in all IRAC bands as well as 2MASS J, H and K bands (not shown here) but not at 24~$\mu$m. It is identified as ``SSTGLMC G305.6517+00.3493'' in the GLIMPSE I Spring '07 Catalog and as ``13145704-6223533'' in the 2MASS All-Sky Point Source Catalog. Based on the 2MASS and IRAC colors of the central source, \citet{Marston2010} have suggested MB3957 as a Wolf-Rayet candidate. The ON position of the IRS slit only partially covers the central source in both the SH and LH modules (see Fig.~\ref{fig:m24_3957}).

The Spitzer/IRS spectrum of MB3957 is presented in Figure \ref{fig:spec3957}. It is dominated by a continuum and several low excitation gas lines whose fluxes are given in Table \ref{tab:flux3957}. At short wavelengths, no PAH features are visible. As shown on the MIPS 24~$\mu$m and IRAC observations (see Fig.~\ref{fig:m24_3957} and \ref{fig:3957_irac}), the background emission is varying significantly around MB3957 and the IRS OFF position is in the lowest surface brightness region. We check that the background surface brightness is not varying enough in the MIPS 24 and IRAC 8~$\mu$m images to account for the residual continuum in the final IRS spectrum of MB3957 and that the MIR continuum is actually arising from the MB. 
In the following, we combine the informations at our disposal (central source photometry and MIR spectroscopy) to constrain the true nature of MB3957.

\paragraph{The MIR gas lines}

The MIR spectrum of MB3957 shows several gas lines listed in Table \ref{tab:flux3957}. These lines are not resolved with IRS ($R\sim600$). We correct the line fluxes for extinction. The details of the correction are given in section \ref{sec:mb3957_wholefit}. The detection of [\ion{Fe}{3}], [\ion{Ne}{2}], [\ion{Ne}{3}], [\ion{S}{3}] and [\ion{S}{4}] lines, along with the non-detection of, for instance, [\ion{Fe}{2}], [\ion{O}{4}] or [\ion{Ne}{5}] lines, sets some constraints on the energy of incident photons. Table \ref{tab:ip} gives the ionization potentials of Iron, Neon, Sulfur and Oxygen for their ionization states that emit in the MIR. There is a range of energy where these elements with these observed ionization states can coexist. Photons with energy ranging from 16 to 41 eV have to be significantly more numerous than those with energy below 16 eV or above 55 eV. Simply based on these ionization potentials, we expect the shell to emit [\ion{O}{3}] 51.8 and 88.4~$\mu$m lines.

\begin{table}[t]
\caption{Ionization potentials of Iron, Neon, Sulfur and Oxygen for their first few ionization states. Values are in bold for mid-IR lines that are detected in MB3957, in italic for mid-IR lines that are not detected, and unchanged for ionization states that do not emit in the mid-IR.}
\begin{center}
\begin{tabular}{l c c c c}
\hline
\hline
Element	& II			& III			& IV			& V \\
\hline
Iron		& 7.9		& \bf{16.19}	& 30.65 \\
Sulfur	& 10.36		& \bf{23.34}	& \bf{34.79}	& 47.22 \\ 
Oxygen	& 13.62		& 35.12		& \it{54.93}	& 77.41 \\
Neon	& \bf{21.56}	& \bf{40.96}	& 63.45		& \it{97.12} \\
\hline
\end{tabular}
\end{center}
\label{tab:ip}
\end{table}%

We first discuss the detection of the [\ion{Fe}{3}]~22.92~$\mu$m line. We start by verifying that it is not the [\ion{Fe}{2}] 22.90~$\mu$m line. The previous discussion on the ionization potentials already suggests that the presence of \ion{Fe}{3} is more likely than that of \ion{Fe}{2}. Then, if it was the [\ion{Fe}{2}] 22.90~$\mu$m line, our models predict we should also see the [\ion{Fe}{2}] 17.9 and 26.0~$\mu$m lines with fluxes at least an order of magnitude higher. Since we do not detect those lines, we conclude that the [\ion{Fe}{2}] 22.90~$\mu$m line is at least a factor 10 fainter than the sensitivity in the LH module which we estimate to be no better than $4\times10^{-22}\ \rm{W.cm^{-2}}$. As a consequence, the [\ion{Fe}{2}] 22.90~$\mu$m line contributes at most 1\% to the feature at 22.9~$\mu$m. Therefore, it is more likely that this line is the [\ion{Fe}{3}]]~22.92~$\mu$m line. This line has rarely been observed: in the supernova remnant RCW103, prototype of a SNR shock heavily interacting with dense ISM \citep{Oliva1998}, in the PNe NGC7009 and NGC7027 \citep{Rubin1997}, and toward the Galactic center \citep{Contini2009}. [\ion{Fe}{3}] lines are more commonly observed in the optical part of the spectrum, even though \citet{Keenan1992} showed the [\ion{Fe}{3}] 22.92~$\mu$m was the strongest line of the Fe $3d^6$ configuration in a 10$^4$~K and 100 to 1000~cm$^{-3}$ gas. Moreover, the [\ion{Fe}{2}] lines at 17.9, 24.5, and 26.0~$\mu$m are usually significantly brighter than the [\ion{Fe}{3}] 22.9~$\mu$m line. In MB3957, the [\ion{Fe}{3}] 22.92~$\mu$m line is at least a factor 10 brighter than the [\ion{Fe}{2}] 24.5 and 26.0~$\mu$m lines. Indeed, we estimate the sensitivity to be no better than $4\times10^{-22}\ \rm{W.cm^{-2}}$ within the LH module, where the [\ion{Fe}{2}] 24.5/26~$\mu$m, [\ion{O}{4}] 25.9~$\mu$m and [\ion{Ne}{5}] 24.3~$\mu$m would be if they were detected.

Then, the detection of an iron line might trace the presence of a shock that processes the dust grains and sends iron atoms back to the gas. In the case of a shock, the sulfur and neon gas lines are in agreement with an electronic density of a few $10^2\ \rm{cm^{-3}}$ \citep{Alexander1999} and a velocity of $200-300\ \rm{km.s^{-1}}$ \citep{Hewitt2009}. The MAPPINGS~III Library of Fast Radiative Shock Models from \citet{Allen2008} also predicts that, within a density of $100\ \rm{cm^{-3}}$, the observed [\ion{Ne}{3}]~15.5/[\ion{Ne}{2}]~12.8, [\ion{S}{3}]~18.7/[\ion{S}{3}]33.5 and [\ion{S}{4}]~10.5/[\ion{S}{3}]33.5 are in agreement with a shock of $200-250\ \rm{km.s^{-1}}$, along with its precursor. However, one would expect to see other MIR iron lines (e.g. [\ion{Fe}{2}] 17.9, 24.5 and 26.0~$\mu$m). Indeed, the fast radiative shock models also predict that the [\ion{Fe}{2}] lines are at least one order of magnitude brighter than [\ion{Fe}{3}]~22.92~$\mu$m. Therefore we suggest that 
photoionization rather than shock is the interpretation for the neon, sulfur and iron lines.

\begin{figure}[!t]
\centering
\subfigure[]
	{\label{fig:SvsNe}
	\includegraphics[angle=90,width=\linewidth]{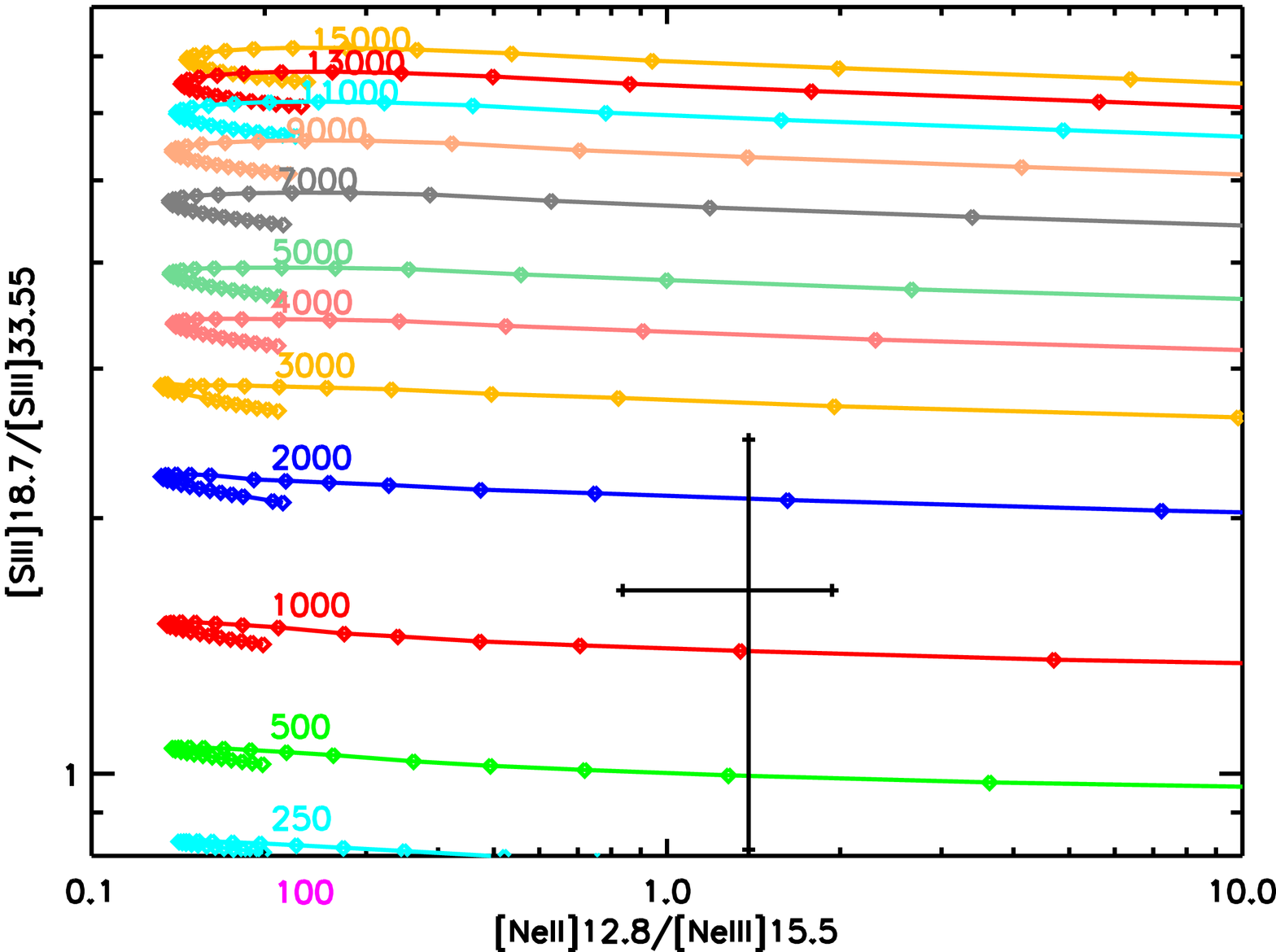}}
\subfigure[]
	{\label{fig:SvsS}
	\includegraphics[angle=90,width=\linewidth]{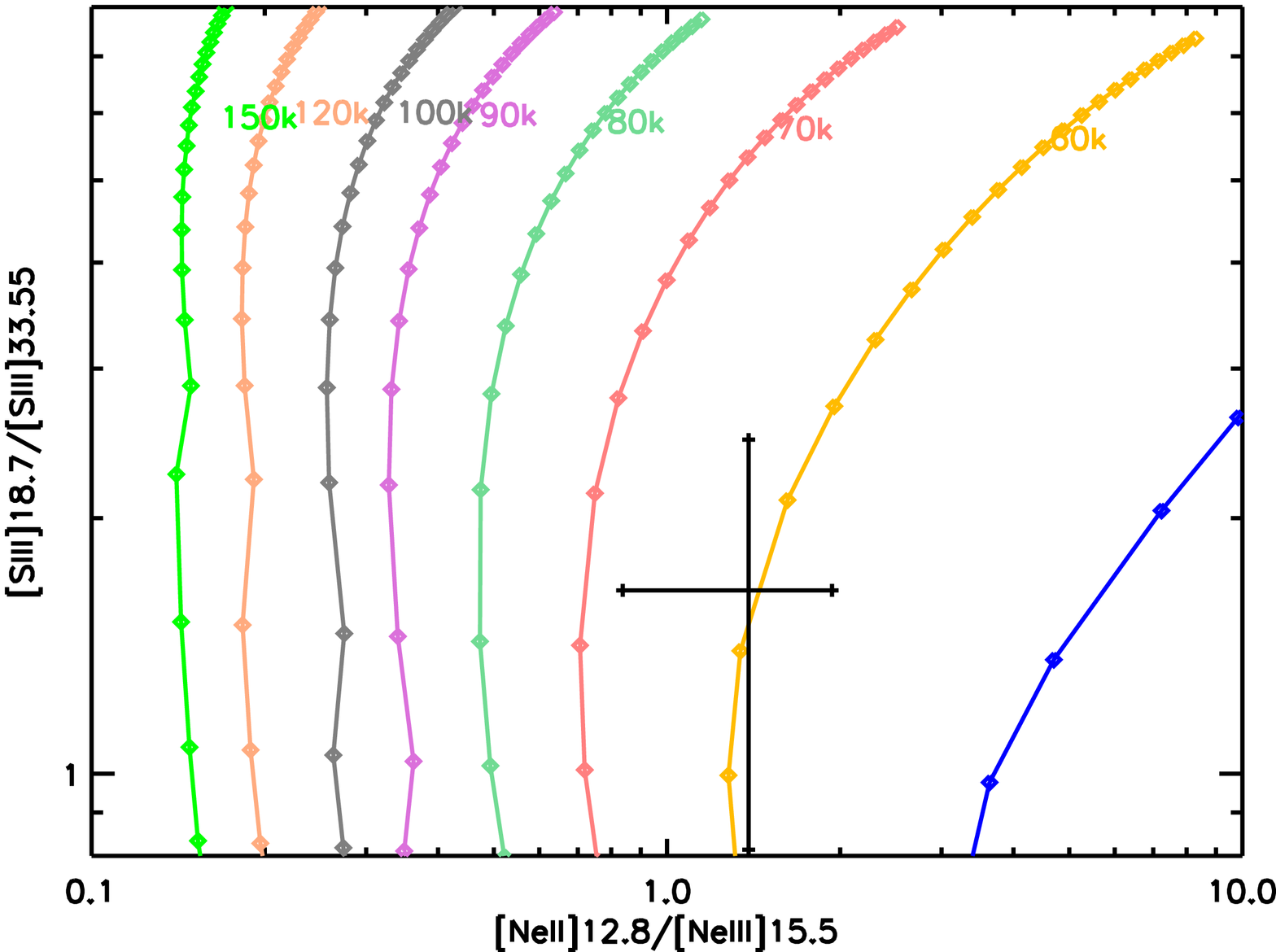}}
\caption{Diagrams of the [\ion{S}{3}]~18.7/[\ion{S}{3}]33.5 ratio as a function of the [\ion{Ne}{2}]~12.8/[\ion{Ne}{3}]~15.5 from our MAPPINGS~III modeling. (a) shows curves for a given gas density from 250 to 15000 cm$^{-3}$. (b) shows curves for a given inner source temperature, from 50000 to 150000 K. The black cross shows the observed ratios for MB3957.}
\label{fig:3957diag}
\end{figure}

We then use the grid of models that we computed with MAPPINGS~III. Figure \ref{fig:3957diag} shows two diagrams of the [\ion{S}{3}]~18.7/[\ion{S}{3}]33.5 ratio as a function of the [\ion{Ne}{2}]~12.8/[\ion{Ne}{3}]~15.5 ratio for a given inner source temperature or a given gas density. The line ratios observed in MB3957 match a central source temperature of 60,000 to 70,000~K, with a gas density of about 250 to 3000~cm$^{-3}$. These results are not significantly altered by the addition of dust in the model. While significantly cooler than the central source of the two dust-poor MBs, the source at the center of MB3957 is significantly hotter than a main sequence star and would match a white dwarf or a Wolf-Rayet star. MAPPINGS~III does not include the [\ion{Fe}{3}] lines fluxes, but it gives the relative abundances and total column densities of the iron first ionization levels. In order to have more \ion{Fe}{3} cations than \ion{Fe}{2} and \ion{Fe}{4} cations, the temperature of the central source has to be between 30,000 and 80,000 K, which is in agreement with the neon and sulfur lines diagnostics while the gas density is not well constrained. The maximum \ion{Fe}{3}-to-\ion{Fe}{2} abundance and column density ratios are both about 3.4 for $(T_{BB}, n_{gas}) = (60000\ K, 3000\ \rm{cm^{-3}})$ and $(50000\ K, 500\ \rm{cm^{-3}})$ respectively, a factor of at least a few smaller than the observed flux ratio between the [\ion{Fe}{3}] 22.9~$\mu$m line and the [\ion{Fe}{2}] 24.5 and 26.0~$\mu$m lines. To explain the remaining discrepancy, we could invoke the difference between the oscillator strength of the [\ion{Fe}{3}] 22.9~$\mu$m line and that of the [\ion{Fe}{2}] 24.5 and 26.0~$\mu$m lines. Unfortunately, these values are not available due to the complexity of the calculations for the \ion{Fe}{3} cation. Difference of extinction between the iron cations emitters, due to layering of photo-ionized matter in the shell could be another interpretation of the remaining discrepancy since \ion{Fe}{2} cations are expected to be farther away from the central source and deeper in the shell structure than the \ion{Fe}{3} cations. However, to attenuate the [\ion{Fe}{2}] lines by a factor 3 larger than the [\ion{Fe}{3}] lines requires an extinction differential of about 1.1 magnitude at wavelengths of about 20~$\mu$m, which corresponds to about 40 magnitudes of visual extinction. This seems unrealistic. The presence of the [\ion{Fe}{3}] 22.9~$\mu$m line without the mid-IR [\ion{Fe}{2}] lines in the spectrum of MB3957's shell remains complicated to interpret due to the difficulty to model with precision the \ion{Fe}{3} cation transitions. However, we suggest the whole set of observed mid-IR gas lines in MB3957 indicate that the shell is produced by the photo-ionization of a few $10^2\ \rm{cm^{-3}}$ gas by an inner source of about $6\times10^4\ \rm{K}$.

\paragraph{The central source}

\begin{table}[b]
\caption{IR magnitudes and fluxes of the source at the center of MB3957.}
\begin{center}
\begin{tabular}{c c c}
\hline
\hline
Band	 & Magnitude			& Fluxes (mJy) \\
\hline
J		& 13.825$\pm$0.034	&  4.704$\pm$0.156\\
H		& 11.884$\pm$0.039	& 18.06$\pm$0.665 \\
K		& 10.674$\pm$0.023	& 35.84$\pm$0.792 \\
3.6~$\mu$m	& 9.618$\pm$0.059	& 39.94$\pm$2.176 \\
4.5~$\mu$m	& 9.041$\pm$0.047	& 43.48$\pm$1.862 \\
5.8~$\mu$m	& 8.732$\pm$0.032	& 36.96$\pm$1.1 \\
8.0~$\mu$m	& 8.490$\pm$0.028	& 25.78$\pm$0.6573 \\
\hline
\end{tabular}
\end{center}
\label{tab:mag_3957}
\end{table}%

\begin{figure*}[!t]
\centering
\subfigure[]
	{\label{fig:fit_cs3957sed_1bb}
	\includegraphics[angle=90,width=.45\linewidth]{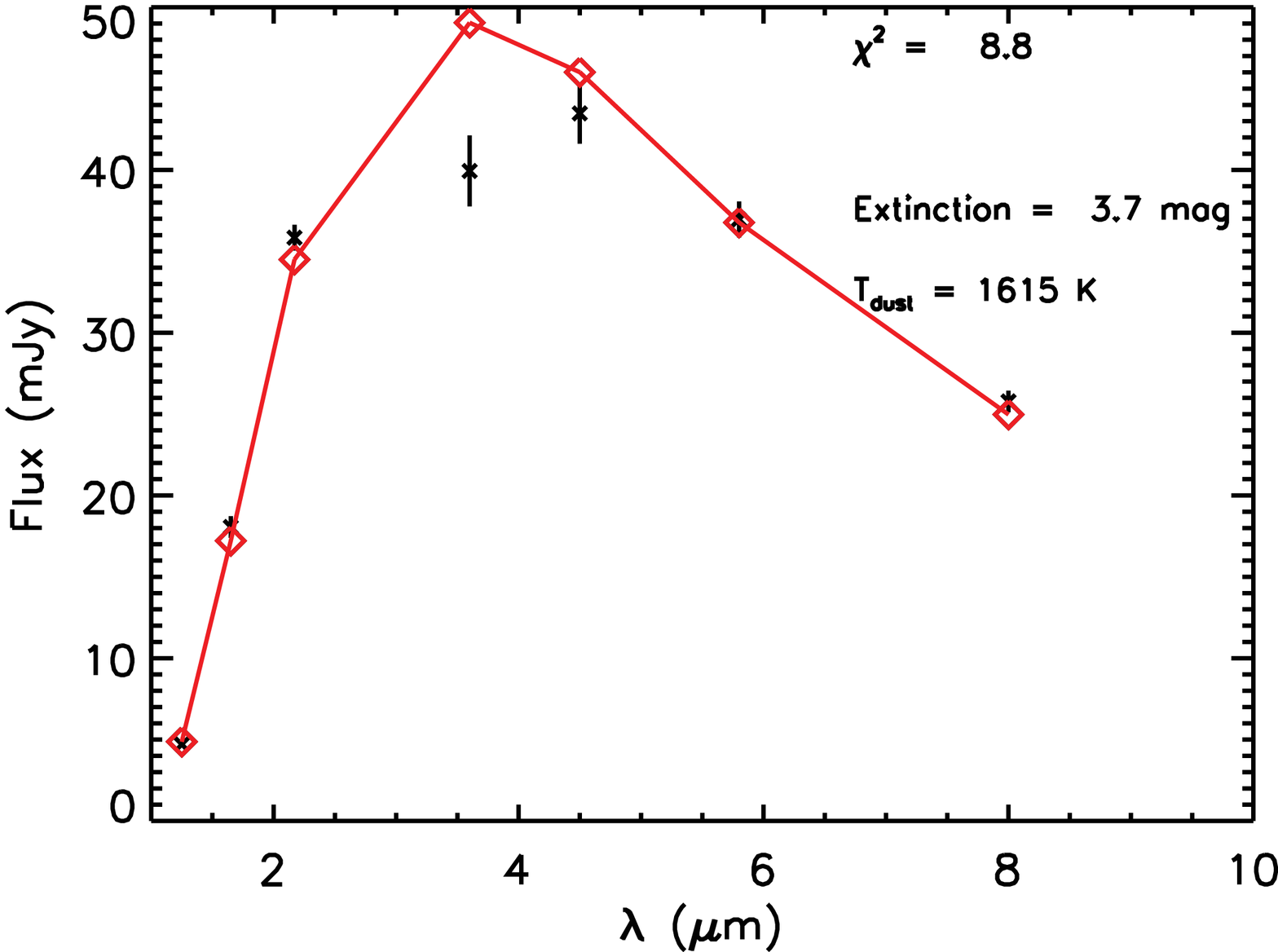}}
\subfigure[]
	{\label{fig:bestfit_cs3957sed_2bb}
	\includegraphics[angle=90,width=.45\linewidth]{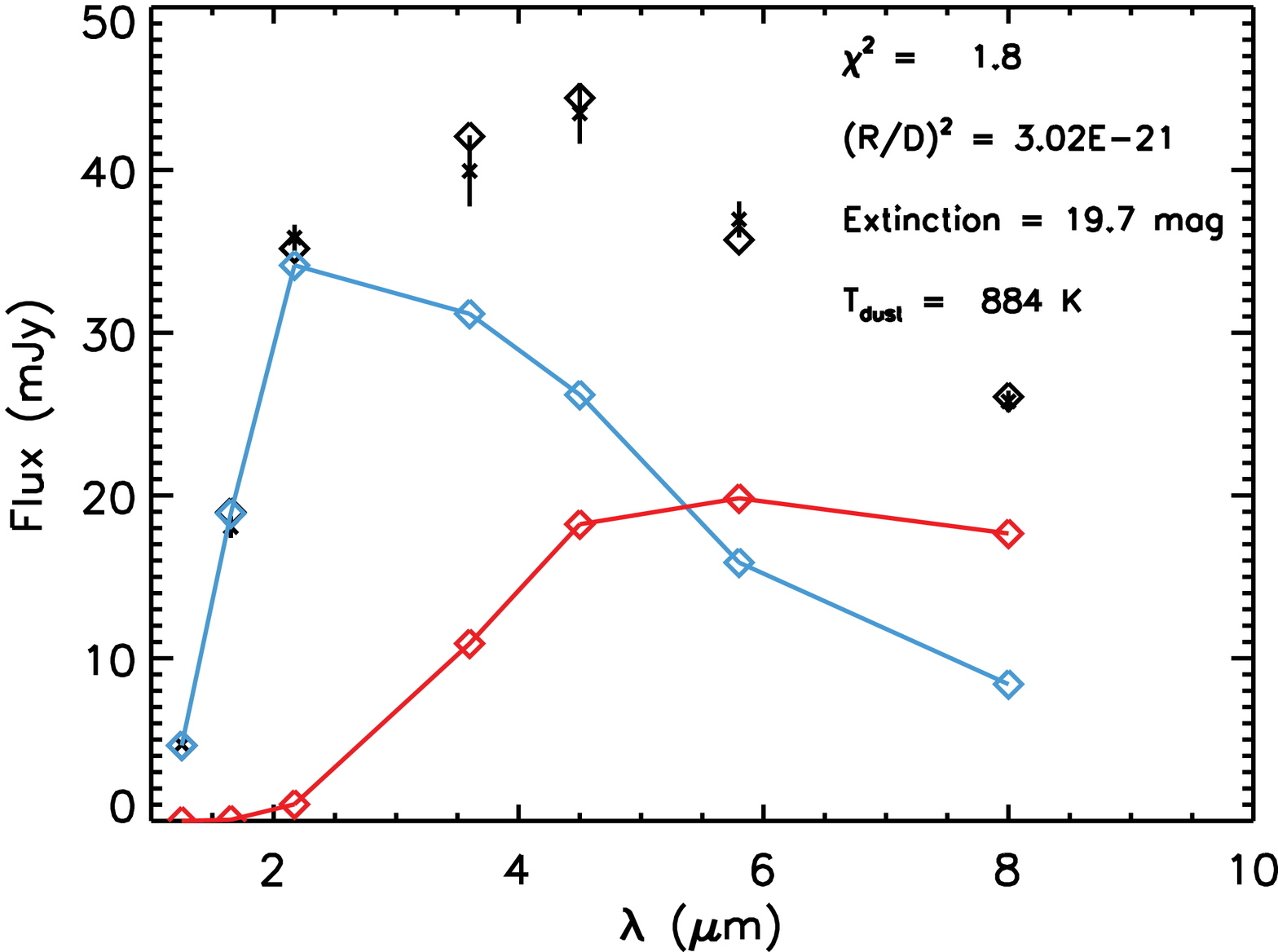}}
\caption{One and two component fitting of MB3957 central source SED. The black crosses are the fluxes from the 2MASS and GLIMPSE catalogs. (a) Best fit with a single component (red diamonds). (b) Best-fit of the central source SED (red diamonds show the cold component, blue diamonds show the hot central source and black diamonds show the total) with two components.}
\label{fig:fit_cs3957sed}
\end{figure*}

The 2MASS and IRAC fluxes of the central source in MB3957 are given in Table \ref{tab:mag_3957} and are shown in Figure \ref{fig:fit_cs3957sed}. The modeling of the MIR gas lines with MAPPINGS~III suggests that the central source has a temperature of about 60000 K, which could fit either a white dwarf or a very hot Wolf-Rayet star, as suggested by \citet{Marston2010}. The SED peaks in the near-IR which clearly indicates a reddened source. Indeed, the best fit of the entire SED with a blackbody leads to a temperature of 1600K (see Fig.~\ref{fig:fit_cs3957sed_1bb}). Reddening might be due to extinction along the line of sight, highly plausible through the Galactic plane. However, with an intrinsic temperature of 60000 K for the source, it's impossible to find a decent fit of the SED, whatever the distance and the extinction. We thus suggest the IR excess in the SED originates in a circumstellar enveloppe of dust or in the presence of a substellar companion around the inner source. The presence of a substellar companion in a binary system that comprises a white dwarf has been reported by \citet{Zuckerman1992}. GD165 and GD1400 are two examples of such systems \citep{Kirkpatrick1999,Farihi2004}. The presence of a disk of dust around a white dwarf has been observed by \citet{Su2007} in the MIPS 24~$\mu$m image of the Helix Nebula. Wolf-Rayet stars have long been known to be surrounded by dust shells. We use the SED of the central source in MB3957 to determine the validity of either interpretation.

We model the central hot source and the cooler source (surrounding disk of dust or brown dwarf companion) with blackbody functions. The best fit depends on the initial guess and usually leads to a one component fit (see Fig.~\ref{fig:fit_cs3957sed_1bb}). The reduced $\chi^2$ for this fit is about 9. In order to ``force'' the use of two components, we fix the dilution factor of the central source $W_{*}=(R_*/d_*)^2$ to non-zero values and run several fits. The other parameters (star temperature, dust temperature, dust scaling factor, extinction scaling factor) are set free with some boundaries: $50000\ \rm{K}<T_{*}<70000\ \rm{K}$, $10\ \rm{K} < T_{dust} < 10000\ \rm{K}$ and $1\ \rm{mag} <A_V< 100\ \rm{mag}$. We use the average NIR extinction values found by \citet{Indebetouw2005} in the Galactic plane. We explore a very wide range of values for $(R_*/d_*)^2$. Small values of $\chi^2$ ($\sim10$) are found for very small values of $(R_*/d_*)^2$, which corresponds to the fit with only one component. However, we find a better fit ($\chi^2\sim2$) for a very narrow range of $(R_*/d_*)^2$ ($2-3\times10^{-21}$). The cool component temperature and the extinction along the line of sight both reach well determined values in that range: $T_{dust}\sim900\ \rm{K}$, $A_V\sim20\ \rm{mag}$. The best fit is shown in Figure \ref{fig:bestfit_cs3957sed_2bb}. The dilution factor for the best fit ($R_*/d_* = (2.5\pm0.5)\times10^{-11}$) allows us to constrain the nature of the central source. If the central source is a white dwarf ($R_*\sim0.01R_\odot$) it is at a distance of $4.5\pm0.5\ \rm{pc}$ which is highly unlikely. Indeed, the source would be one of the closest star to Earth. Moreover, the best fit extinction thus corresponds to a column density of $3.7\times10^{22}\ \rm{cm^{-2}}$, using $A_V/N_H=0.53\times10^{-21}\ \rm{cm^{-2}}$ from \citet{Savage1979}. Therefore, the average density along the line of sight would be about $3\times10^3\ \rm{cm^{-3}}$, several order of magnitude larger than that of the diffuse medium. If the central source is a WR star ($R_*\sim10R_\odot$) it is at a distance of $4.5\pm0.5\ \rm{kpc}$ which is much more likely as it corresponds to Galactic distances. Moreover, the average density would be about $3\ \rm{cm^{-3}}$ which is very similar to that of the Galactic diffuse ISM. We thus conclude that the central source of MB3957 is a WR star and that the reddening of its SED is due to the presence of an inner circumstellar enveloppe of hot dust. Our conclusion therefore confirms the interpretation from \citet{Marston2010}. We characterize the dust component in the following section.

\paragraph{Full SED fitting}
\label{sec:mb3957_wholefit}

\begin{figure*}[!t]
 	\centering
 	\includegraphics[angle=90,width=\linewidth]{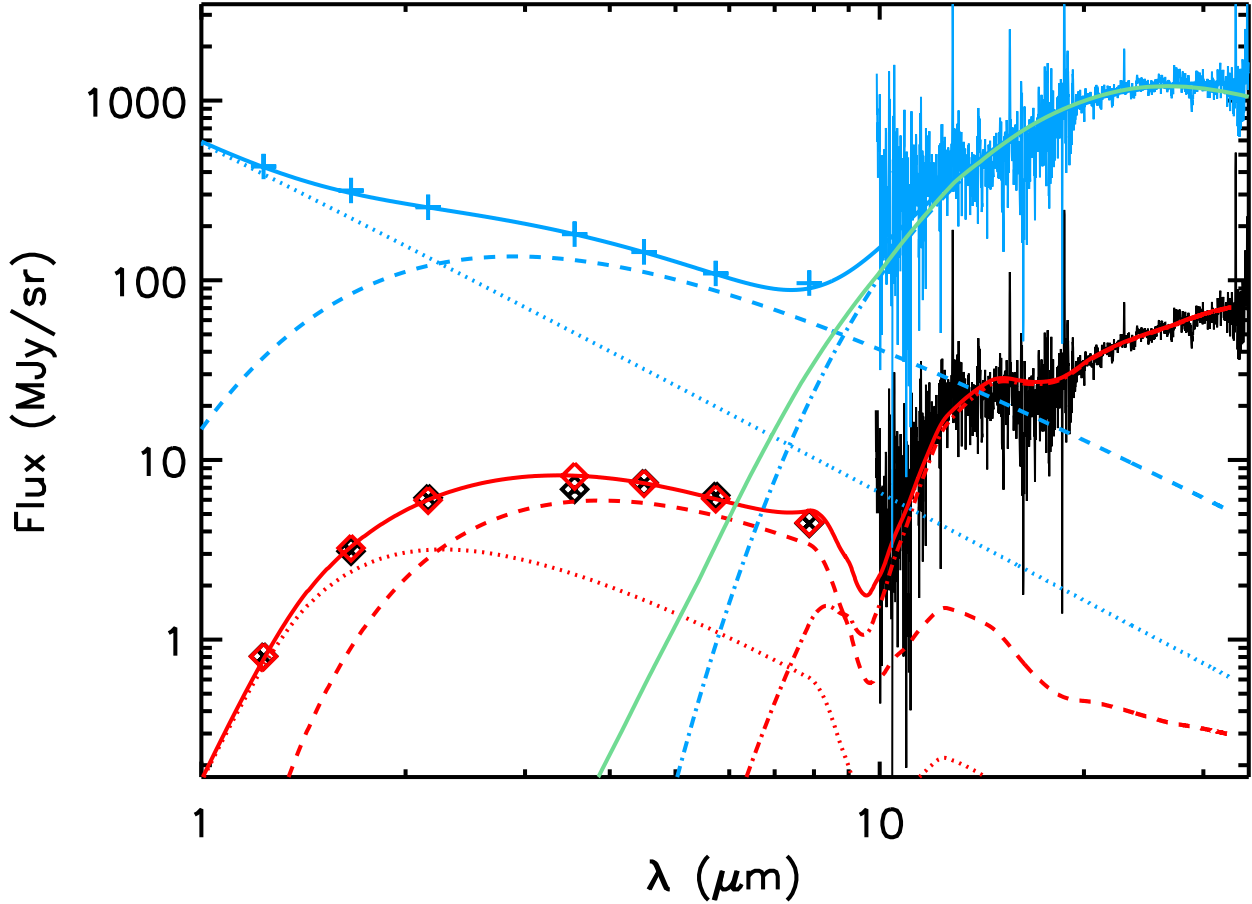}
 	\caption{Three blackbody component fitting of the whole SED of MB3957.
 MB3957's whole SED (black line and diamonds), best fit (red lines and diamonds), dereddened SED multiplied by a factor 10 (blue lines and crosses) and small dust grains model multiplied by a factor 10 (solid green line).}
 	\label{fig:fit_cs3957sed_3bb_abs}
 \end{figure*}

The complete SED of MB3957, combining the 2MASS+IRAC SED and the IRS spectrum, is shown in Figure \ref{fig:fit_cs3957sed_3bb_abs} with an intensity scale that allows a better view on the continuum emission. The 2MASS and IRAC observations are well interpreted by the two component fitting presented in the previous section. In the IRS wavelength range, the emission is dominated by a continuum that shows silicate absorption features at 10 and 20~$\mu$m and that dominates the emission in the MIPS 24~$\mu$m. Therefore, we believe the entire SED of MB3957 can be interpreted with three emission components: the central source, an inner disk of hot dust and an outer shell of warm dust.

In order to confirm this interpretation, we model the entire IR SED of the shell and the central source with three blackbodies attenuated by extinction along the line of sight. We combine the so-called ``ISM'' and ``Galactic center'' extinction curves from \citet{Chiar2006} into several extinction curves, defined by the relative contribution of the ``ISM'' extinction curve ($f_{ISM}$). We set free seven parameters: the temperatures ($T_{*}, T_{hot}, T_{warm}$) and dilution factors ($W_{*}, W_{hot}, W_{warm}$) of the blackbodies, and the normalization of the extinction curve at 9.7~$\mu$m ($\tau_{9.7}$). We put limits on the three temperatures: 50000 to 70000 K for the central source, 500 to 2000 K for the hot dust and 10 to 200 K for the warm dust. We then run a grid of fits with the parameters $W_{*}$ and $f_{ISM}$ fixed at pre-defined values: $f_{ISM}$ varies between 0 and 1 and $W_{*}$ varies from $10^{-22}$ to $10^{-20}$, following the results of previous section. The lower values of $\chi^2$ are obtained for $f_{ISM} \sim 0$ and for $W_{*} \sim 0.9\times10^{-21}$, which is only a factor of a few smaller than the value found for the fit of the 2MASS and IRAC SED only. As a consequence, the conclusion on the nature of the central source remains the same: MB3957 is a WR. The distance to MB3957 is $7.5\times(R_*/10 R_\odot)$ kpc. We do not show here the contour plot for the other parameters but we discuss their values and variations. For the best fit, the depth at 9.7~$\mu$m and the temperatures of the hot and warm component are 2.0, 1750 K and 167 K respectively. The warm component temperature does not vary significantly within the grid we explore, from 165 to 185 K, as it is well constrained by the IRS spectrum. The hot component temperature increases from 500 to 2000 K as $W_{*}$ decreases but is noticeably above 1250 K for $W_{*} < 2\times10^{-21}$. Those values are in agreement with previous observations of WR stars \citep[e.g.][]{vanderHucht1996}. At those temperatures, silicates would begin to be detected in emission rather than in absorption except in the case of carbon rich environment, from which silicates would be absent. The depth $\tau_{9.7}$ decreases from 4 to 1.5 as $W_{*}$ decreases but is noticeably below 2.5 for $W_{*} < 2\times10^{-21}$. Since $f_{ISM} \sim 0$  for the best fit, we use the ``Galactic center'' extinction curve from \citet{Chiar2006} to convert $\tau_{9.7}=2.0$ to $A_V=13.7$ mag, which is in good agreement with the extinction obtained in the previous section.
The temperature of the central source spans the entire range of allowed values, varies quickly from 50000 to 70000 K as $W_*$ decreases from $10^{-21}$ to $3\times10^{-22}$, and for the best fit $T_* =  51000$ K. All parameters are significantly more dependent on $W_{*}$ than $f_{ISM}$.

The best fit is shown in Figure \ref{fig:fit_cs3957sed_3bb_abs} along with the dereddened spectrum of MB3957. The warm component is by far the main contributor of the emission within the IRS wavelength range while the NIR data points can almost only be accounted for by the central source and the hot component. This explains why the results from the fit of the IRAC and 2MASS data points only are not significantly different.The shape of the continuum in the IRS range is almost perfectly reproduced by the fit even though one may argue the silicate feature at 20~$\mu$m seems deeper in the observations. However, taking into account the noise in the IRS/SH data, and the possible variations of the silicate features in the extinction curve not accounted for by the curves of \citet{Chiar2006}, we believe the three blackbody components model is a simple but efficient way to interpret the continuum emission of MB3957. The dereddened curve, also shown in Figure \ref{fig:fit_cs3957sed_3bb_abs}, presents the intrinsic emission spectrum of MB3957 without correcting for the dilution factor and scaled up by a factor 10 for more clarity. The discrepancy at the very short end of the IRS wavelength range might also be explained by a lower S/N ratio combined with a higher extinction correction factor.

\paragraph{Dust mass loss}

Here we finally verify that the distance to MB3957, assumed to be 7.5 kpc, its MIPS 24~$\mu$m brightness and the fit of its whole SED are consistent with the WR interpretation. We model the warm component, which accounts for the emission of dust in the MIPS 24~$\mu$m shell of MB3957, with a more realistic emissivity function. We use the dust model from \citet{Compiegne2011}. For the incident radiation field, we use a 50000 K blackbody at a distance of 12'' to the peak emission of the shell, as given by the MIPS 24~$\mu$m image and which corresponds to 0.4 pc at a distance of 7.5 kpc from the Sun. We plot the emission spectrum of the only small amorphous carbon (SamC) grains in Figure \ref{fig:fit_cs3957sed_3bb_abs}, scaled to match the IRS spectrum. The required dust column density is $2.7\times10^{-8}\ \rm{g.cm^{-2}}$ if we use SamC only. For large amorphous carbon (LamC) or amorphous silicate (aSil) grains, the required dust column density is $\sim1.3\times10^{-7}\ \rm{g.cm^{-2}}$ but the spectral shape of the IRS spectrum is not reproduced as well as with SamC. This might be an indication that large dust grains are ground into smaller one, though a more thorough modeling of the dust size distribution is required. Integrated over the MB3957, the total mass of the shell is $10^{-3} \times (0.01/x_d) \times (d/7.5{\rm kpc})^2\ \rm{M_\odot}$ where $x_d$ is the dust-to-gas mass ratio. Therefore, the MIPS 24~$\mu$m shell of MB3957 would have been formed in $t = 2125 \times d_*({\rm kpc})/(v_{wind}/{\rm 500\ km.s^{-1}}))$ years where $v_{wind}$ is the average stellar wind speed that produced the shell of the WR star. We use a reference velocity of ${\rm 500\ km.s^{-1}}$ since the mid-IR lines are not resolved with the resolution of IRS ($R\sim600$). The formation timescale is about 2 orders of magnitude shorter than the mean lifetime of a WR \citep[$5\times10^5\ \rm{yr}$,][]{Maeder1987}, which allows for multiple mass loss events. The MIPS 24~$\mu$m shell thus traces a very specific time in the WR's life. We then infer the average mass loss rate of MB3957 to be $4.8 \times 10^{-6} \times (0.01/x_d) \times (d/7.5{\rm kpc}) \times (v_{wind}/{\rm 500\ km.s^{-1}})\ \rm{M_\odot/yr}$, a factor of a few lower than the $10^{-5}$ to $10^{-4}\ \rm{M_\odot.yr^{-1}}$ that WR stars have long been known to eject through powerful winds \citep[e.g.,][]{Willis1991}. However, the shell mass, timescale and wind velocity we obtain here are estimates within a factor of a few at least. For instance, \citet{Sahai2000} have shown that the dust-to-gas mass ratio in the planetary nebula with WR type central star He 2-113 is only 0.5\%. Typical values of 0.25\% to 1\% are found in the circumstellar envelopes of AGB stars, which leads to a factor 4 uncertainty. Moreover, we assume the hot and warm component are blackbodies to estimate the distance to MB3957. If we model both with a $\nu^2B_\nu$ function instead, the best fit (not shown here) is found for slightly different parameters. The best fit dilution factor of the star, for instance, corresponds to $11\times10^{-21}$. It still implies that MB3957 is a WR star and places it at a distance of $2.1\times(R_*/10 R_\odot)$ kpc, more than a factor 3 closer than the 7.5 kpc determined from the three blackbodies fitting.

Combining all the observational constraints on MB3957, from near- to mid-IR broadband photometry and mid-IR spectroscopy, we conclude that the most likely interpretation for this shell is that it is a WR surrounded by two circumstellar envelopes. The inner one, visible as an excess in the IRAC bands, is not resolved and may be related to a recent mass loss. The larger one, visible in the MIPS 24~$\mu$m image, is related to an older mass loss. The modeling of the outer shell torus-like morphology which has been related to the presence of a companion in binary systems \citep{Morris1981,Livio1993}, is beyond the scope of this paper and would benefit from better angular resolution.

\subsubsection{Dust-rich object MB4121}
\label{lab:mb4121}

\begin{figure}[!t]
	\centering
	\includegraphics[width=\linewidth]{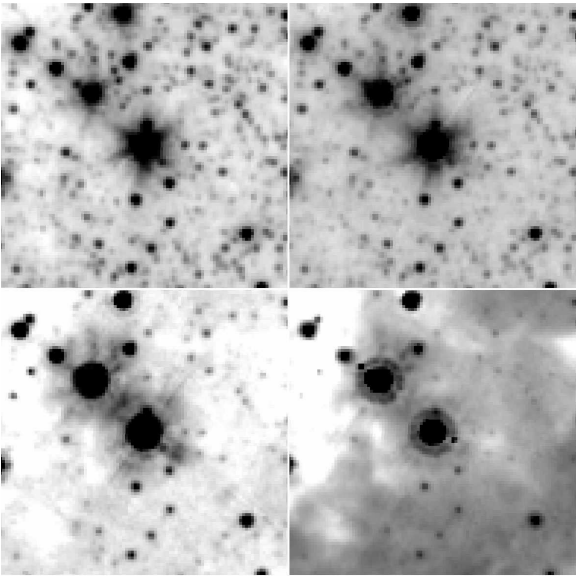}
	\caption{IRAC 3.6 (top-left), 4.5 (top-right), 5.8 (bottom-left) and 8.0~$\mu$m (bottom-right) images (2\arcmin x 2\arcmin) of MB4121.}
	\label{fig:4121_irac}
\end{figure}

The morphology of MB4121 is that of a shell with a central source in the MIPS 24~$\mu$m image (see Fig.~\ref{fig:m24_4121}). At shorter wavelengths (IRAC and 2MASS observations), only the central source is visible (see Fig.~\ref{fig:4121_irac}). What looks like diffuse emission on the IRAC 5.8~$\mu$m is related to instrumental artifact due to the bright central source. A nearby IR dark cloud (IRDC) is visible in the IRAC and MIPS 24~$\mu$m bands. The central source is identified as ``SSTGLMC G337.5543+00.2198'' in the GLIMPSE I Spring '07 Catalog and as ``16364278-4656207'' in the 2MASS All-Sky Point Source Catalog. The IR magnitudes and fluxes for the star are reported in Table \ref{tab:mag_4121}. \citet{Wachter2010} have recently identified the central source of MB4121 as a Be/B[e]/LBV thanks to the striking similarity of its near-IR spectrum with that of G024.73+00.69.

\begin{figure*}[!t]
\centering
\subfigure[]
	{\label{fig:4121_irs_cs}
	\includegraphics[angle=90,width=.45\linewidth]{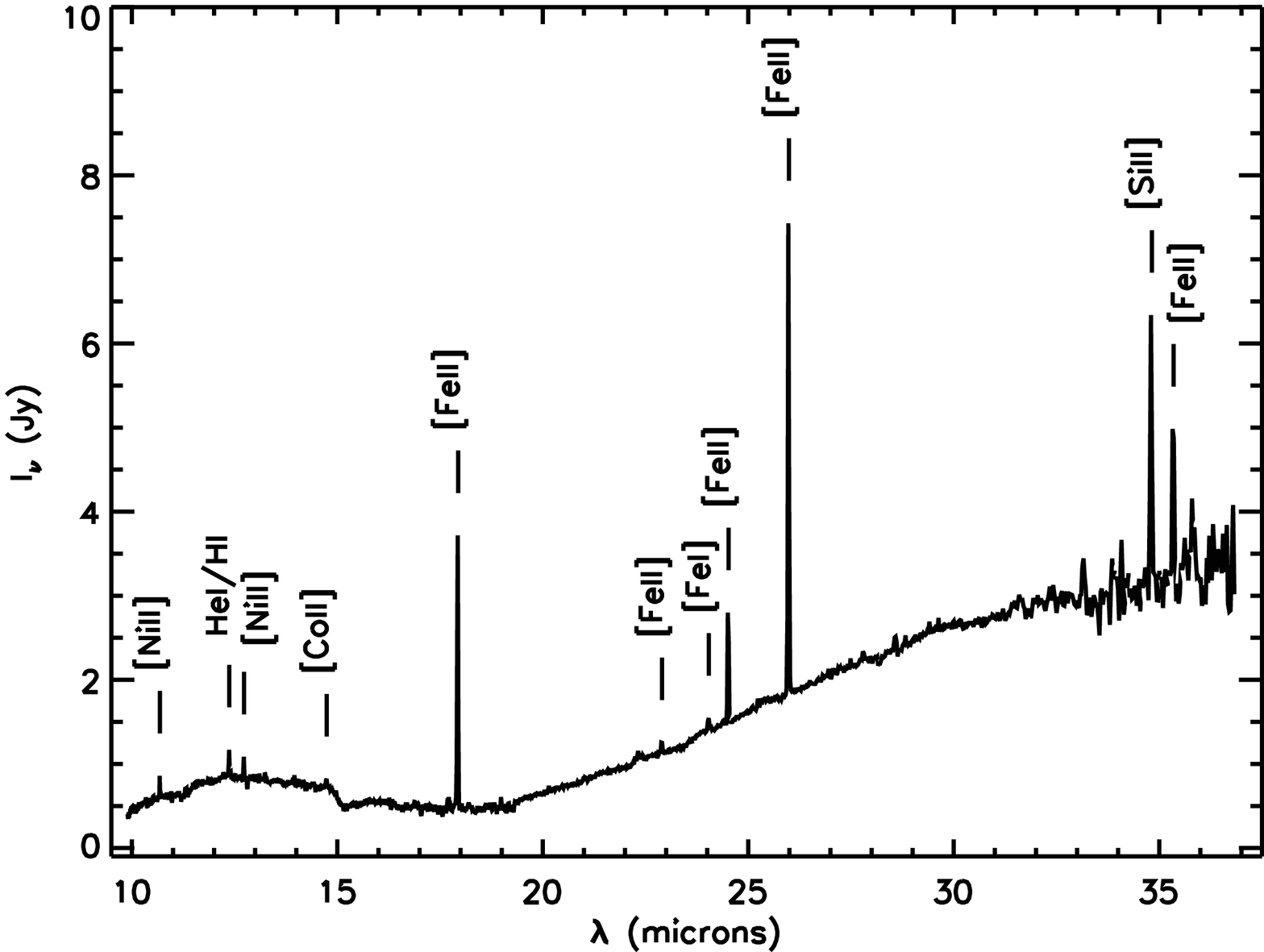}}
\subfigure[]
	{\label{fig:4121_irs_shell}
	\includegraphics[angle=90,width=.45\linewidth]{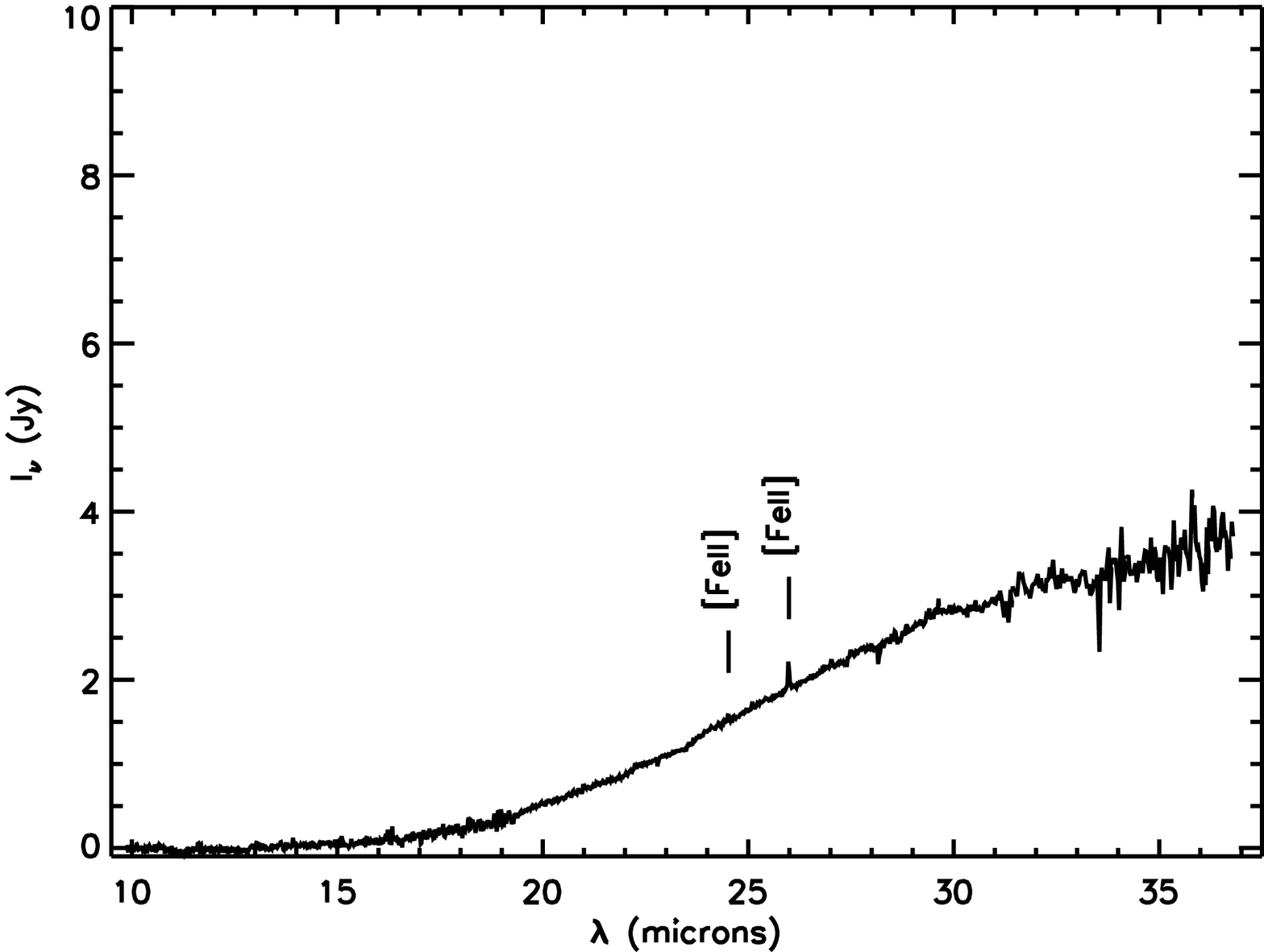}}
\caption{(a) IRS spectrum toward the central source of MB4121. The main gas lines are indicated. (b) IRS spectrum toward the shell's rim of MB4121. The [\ion{Fe}{2}] 24.52~$\mu$m line is detected in only one nod.}
\label{fig:4121_irs}
\end{figure*}

Contrary to the previous three objects presented in this paper, we observe MB4121 toward two positions. The first position is on the central source of the shell and the second is on the northern rim of the shell (see Fig.~\ref{fig:m24_4121}). The two background subtracted IRS spectra are presented in Figure \ref{fig:4121_irs}. A long wavelength ($\lambda > 20 \mu$m) continuum is detected toward both positions. Therefore we attribute this emission component to the outer shell. The spectrum toward the central source shows a richer spectrum with a short wavelength ($\lambda < 20 \mu$m) continuum, absorption features of silicates and CO$_2$ ,and several emission gas lines, most of them from iron (see Fig.~\ref{fig:4121_irs_shell}). The measured fluxes of the main gas lines labeled in the IRS spectrum are given in Table \ref{tab:flux4121a} and \ref{tab:flux4121b}.

We first correct for extinction the IRS spectra using the whole SED of the central source in MB4121. We then use the MIR gas lines as a diagnostic to constrain the temperature of the central source and the density in the object. We finally discuss the outer shell continuum emission.

\begin{table}[b]
\caption{IR magnitudes and fluxes of the source at the center of object \#4121.}
\begin{center}
\begin{tabular}{c c c}
\hline
\hline
Band	 & Magnitude			& Fluxes (mJy) \\
\hline
J			& 11.711$\pm$0.022	& 32.97$\pm$0.67\\
H			& 7.889$\pm$0.042	& 715.7$\pm$27.7 \\
K			& 5.82$\pm$0.021		& 3133$\pm$61 \\
3.6~$\mu$m	& 4.915$\pm$0.041	& 3039$\pm$114 \\
4.5~$\mu$m	& 4.519$\pm$0.034	& 2798$\pm$89 \\
5.8~$\mu$m	& 4.046$\pm$0.017	& 2768$\pm$43 \\
8.0~$\mu$m	& 4.054$\pm$0.015	& 1532$\pm$22 \\
\hline
\end{tabular}
\end{center}
\label{tab:mag_4121}
\end{table}

\begin{table}[b]
\caption[ ]{\label{tab:flux4121a} Lines fluxes detected in the IRS spectrum of MB4121 toward the central source. Units are $10^{-20}\ \rm{W.cm^{-2}}$. The extinction corrected line fluxes results from the whole SED fitting (see section \ref{sec:mb3957_wholefit}).}
\begin{center}
\begin{tabular}{l r@{.}l@{$\pm$}r r@{.}l@{$\pm$}l}
\hline
\hline
Line & \multicolumn{3}{c}{Flux} & \multicolumn{3}{c}{Corrected flux} \\
\hline
~[\ion{Ni}{2}]~10.68$\mu$m &  0&9 &  0.4 &  5&0 &  2.0 \\
~[\ion{He}{1}] or [\ion{H}{1}]~12.37$\mu$m &  1&5 &  0.7 &  4&0 &  2.0 \\
~[\ion{Ni}{2}]~12.73$\mu$m &  1&1 &  0.4 &  3&0 &  1.0 \\
~[\ion{Co}{2}]~14.74$\mu$m &  0&3 &  0.5 &  0&8 &  1.5 \\
~[\ion{Fe}{2}]~17.94$\mu$m &  9&0 &  0.6 & 35&0 &  2.0 \\
~[\ion{Fe}{2}]~22.90$\mu$m &  0&3 &  0.7 &  0&9 &  2.0 \\
~[\ion{Fe}{1}]~24.04$\mu$m &  0&3 &  0.3 &  1&0 &  0.8 \\
~[\ion{Fe}{2}]~24.52$\mu$m &  2&6 &  0.3 &  7&3 &  0.8 \\
~[\ion{Fe}{2}]~25.99$\mu$m &  8&4 &  0.3 & 22&0 &  0.8 \\
~[\ion{Si}{2}]~34.82$\mu$m &  4&0 &  1.0 &  7&0 &  2.0 \\
~[\ion{Fe}{2}]~35.35$\mu$m &  3&0 &  2.0 &  4&0 &  3.0 \\
\hline
\end{tabular}
\end{center}
\end{table}

\begin{table}[b]
\caption[ ]{\label{tab:flux4121b} Gas lines detected in the IRS spectrum of MB4121 toward the shell. Units are $10^{-20}\ \rm{W.cm^{-2}}$. We do not correct those fluxes for extinction.}
\begin{center}
\begin{tabular}{l r r}
\hline
\hline
Line & Flux & Uncertainty \\
\hline
~[\ion{Fe}{2}]~24.52~$\mu$m	&  0.1 & 0.5 \\
~[\ion{Fe}{2}]~25.99~$\mu$m	&  0.6 & 0.3 \\
\hline
\end{tabular}
\end{center}
\end{table}

\paragraph{Extinction correction}
\label{lab:4121_fullfit}

The spectrum toward the central source shows absorption features from silicates at 9 and 18~$\mu$m and CO$_2$ at 15~$\mu$m (see figure \ref{fig:4121_irs_cs}). These features may arise from the IRDC visible in figures \ref{fig:rgb_4121} and \ref{fig:4121_irac} at a position close to that of MB4121. The outer shell of MB4121 may also contribute to the extinction along the line of sight toward the central source. We correct the emission spectrum of MB4121 to take into account the extinction.

\begin{figure*}[!t]
\centering
\subfigure[]
	{\label{fig:4121_co2_abs}
	\includegraphics[angle=90,width=.45\linewidth]{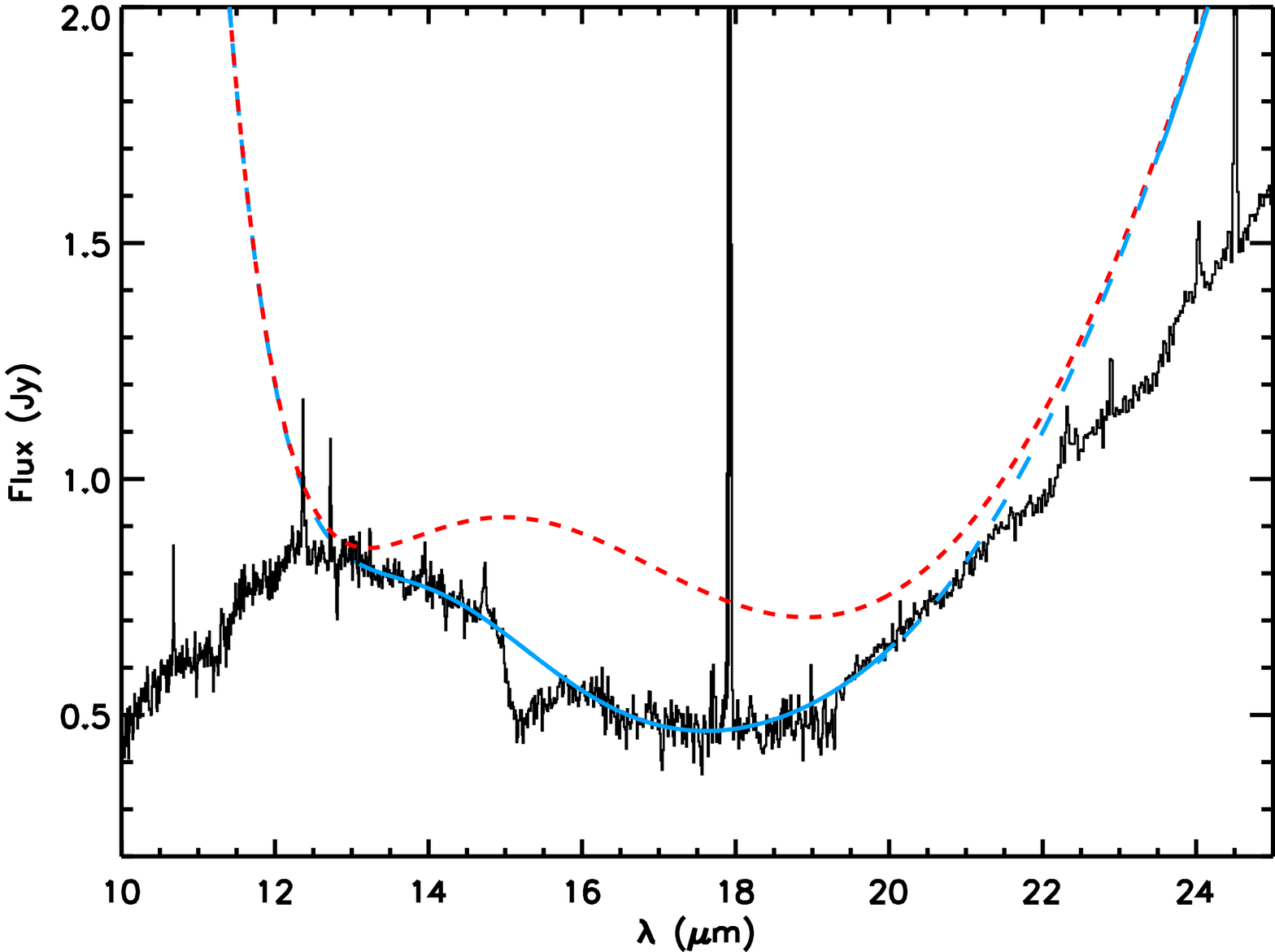}}
\subfigure[]
	{\label{fig:4121_co2_od}
	\includegraphics[angle=90,width=.45\linewidth]{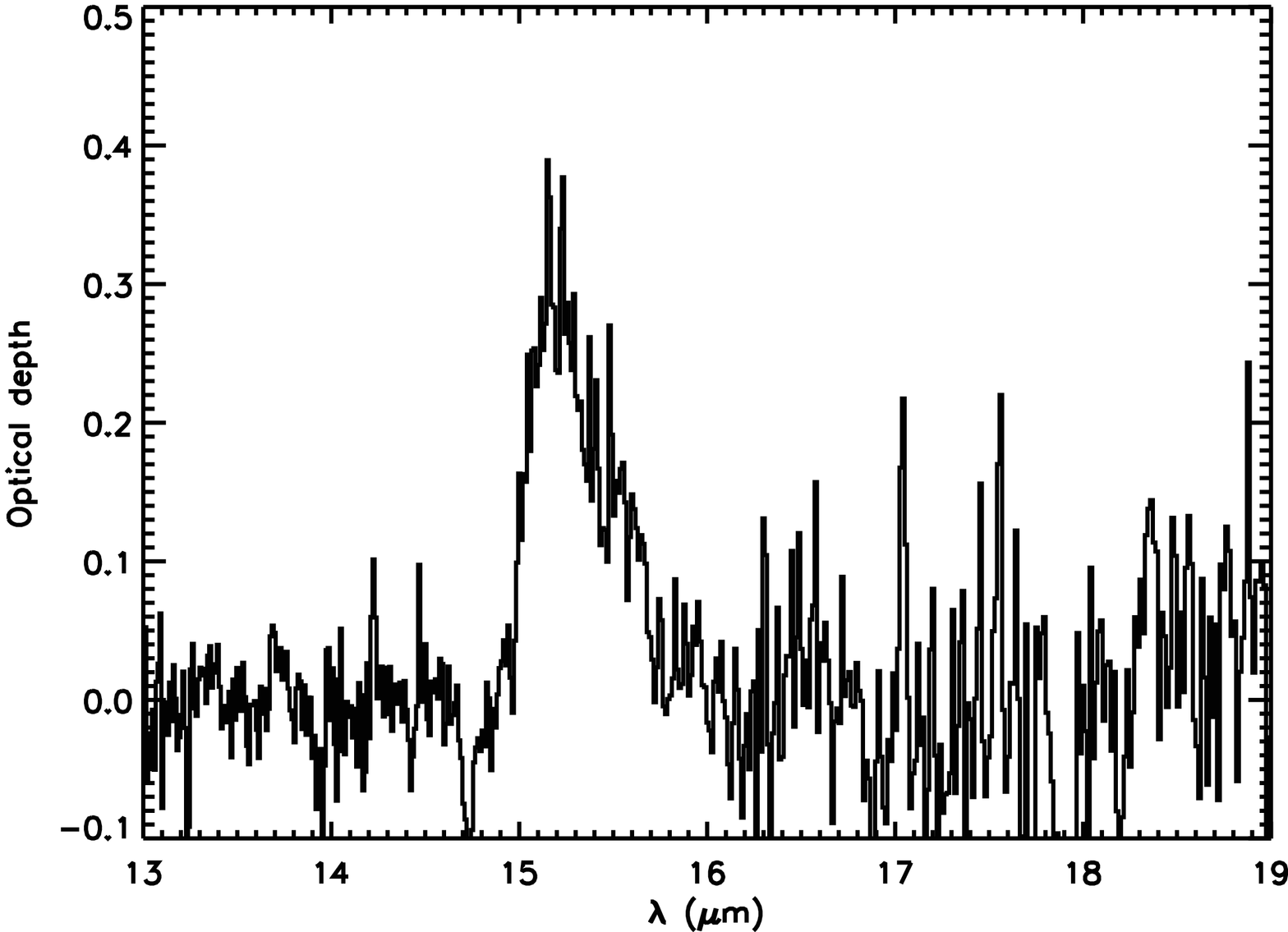}}
\caption{(a) The black solid line is the IRS spectrum toward the central source of MB4121. The red dashed line is the polynomial fit to the continuum and the solid blue line is the best fit of the continuum with a Gaussian included to simulate the blue wing of the 18~$\mu$m silicate feature, following the method presented in \citet{Pontoppidan2008}. (b) Optical depth of the CO$_2$ bending mode.}
\label{fig:4121_co2}
\end{figure*}

First, we extract the profile of the CO$_2$ bending mode at 15~$\mu$m following the method presented in \citet{Pontoppidan2008}. The continuum is fitted by a third order polynomial function in the wavelength ranges from 13 to 14.7~$\mu$m, from 16.2 to 17.5~$\mu$m and from 18.5 to 21~$\mu$m in order to avoid the emission gas lines of [\ion{Co}{2}]~14.74~$\mu$m and [\ion{Fe}{2}] 17.94~$\mu$m. A Gaussian is included to simulate the blue wing of the 18~$\mu$m silicate feature. Figure \ref{fig:4121_co2_abs} illustrate the method and figure \ref{fig:4121_co2_od} shows the resulting optical depth profile of the CO$_2$ bending mode. We estimate the corresponding column density of CO$_2$, $N_{CO_2} = A_{CO_2}^{-1}\times \int\tau d\nu = 5.7\times10^{17}\ \rm{cm^{-2}}$, where the absorbance $A_{CO_2} = 1.1\times10^{-17}\ \rm{cm}$ \citep{Gerakines1995}. To convert this CO$_2$ column density into magnitudes of extinction, one has to know the abundance of CO$_2$ relative to hydrogen. \citet{Whittet2007} found a tight correlation between $A_V$ and $N_{CO_2}$ in the Taurus quiescent intracloud medium, $N_{CO_2} = q(A_V-A_0)$ where $q=(0.252\pm0.036)\times10^{17}\ \rm{cm^{-2}}$ and $A_0 = 4.3\pm1.0$ mag which leads to $A_V = 27\pm4$ mag toward MB4121. However, one shall consider this an upper-limit on the visual extinction as the IRDC, which may account for most of the CO$_2$ absorption is not representative of the matter along the whole line of sight toward MB4121. We remove the CO$_2$ absorption feature by interpolating the fit of the continuum between 14.8 and 16.4~$\mu$m.

\begin{figure}[!t]
	\centering
	\includegraphics[angle=90,width=\linewidth]{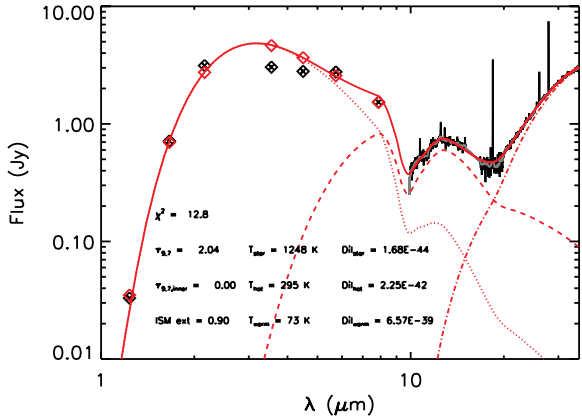}
	\caption{Whole IR SED of the central source in MB4121. Black diamonds are the 2MASS and IRAC data, black solid line is the IRS spectrum. The best fit model is in red: dotted and dashed lines are the three $\nu^2B\nu(T)$ components, the solid line is the total, the diamonds are the transmitted fluxes in the 2MASS and IRAC data.}
	\label{fig:4121_best_fit}
\end{figure}

Then, our goal is to correct for the silicate extinction using the whole SED of the central source in MB4121 in a way similar to that used for MB3957 (see section \ref{sec:mb3957_wholefit}). However, in those conditions, the fit of MB4121 does not lead to a satisfying result: the stellar component fits the 2MASS and IRAC data but the required extinction is too large and the distance too close. We thus proceed by step to narrow down the number and temperatures of the emission component. First, we fit the IRAC and 2MASS observations with a single component and extinction. If we use a blackbody, then the best fit corresponds to $T_*\sim7000$ K, $W_*\sim6\times10^{-18}$, and  $\tau_{9.7}\sim4$ or $A_V\sim30$ mag. Assuming the star is a very cool B-star, with a radius $R_* = 2 R_\odot$, its distance is less than 20 pc from the Sun, which seems unlikely. Moreover, the extinction along the line of sight leads to an average density of more than $250\ \rm{cm^{-3}}$, about two orders of magnitude higher than the diffuse ISM. Therefore, we conclude the IRAC and 2MASS observations reddening is not only due to extinction but also to a dusty component. We then assume the IRAC and 2MASS observations trace an inner envelope of hot dust. We use a $\nu^2B_\nu(T)$ function to fit the IRAC and 2MASS data. The best fit is found for $T_{hot}\sim950$ K and $A_V\sim10$ mag. On the other end of the spectrum, we also use a $\nu^2B_\nu(T)$ function to fit the IRS spectrum for wavelengths $\lambda >19 \mu$m, to constrain the properties of the continuum originating in the outer shell. The best fit corresponds to $T_{cool}\sim80$ K. Finally, we combine the IRAC, 2MASS and IRS observations of the central source and fit them with three graybody ($\nu^2B_\nu(T)$) components: a hot envelope of hot dust ($T_{hot}$) that accounts for the 2MASS and IRAC observations, a cool outer shell ($T_{cool}$) that is detected in the MIPS 24~$\mu$m images and an warm inner component ($T_{warm}$) that is necessary to account for the emission between 10 and 20~$\mu$m and that could be related to an ejecta. We use our preliminary fit of the short and long wavelengths SED to put limits on the components temperatures: $500<T_{hot}<1500$ K, $50<T_{cool}<100$ K and $100<T_{warm}<500$ K. We run a grid of fit for different values of $f_{ISM}$, the contribution of the ``ISM'' extinction curve to the extinction along the line of sight, between 0 and 1. The resulting $\chi^2$ depends only slightly on $f_{ISM}$. The best fit corresponds to $T_{hot}=1250$ K, $T_{cool}=74$ K and $T_{warm}=296$ K and 19 magnitudes of visual extinction along the line of sight. The best fit is shown in Figure \ref{fig:4121_best_fit}. The agreement with the observations is very good, except for the IRAC 3.6 and 4.5~$\mu$m data points, but these measurements might be affected by saturation, as the associated flags in the ``less reliable'' GLIMPSE catalog suggest. We use the extinction curve derived from this fit to correct the MIR gas line fluxes (see Table \ref{tab:flux4121a}).

The visual extinction that we derive from the IR SED fitting (19 mag) is in agreement with the upper-limit we derive from the CO$_2$ absorption feature (27 mag). We obtain a third estimate of the extinction thanks to the IRAC 8~$\mu$m and MIPS 24~$\mu$m surface brightness of the IRDC that can be seen near MB4121. We extract several profiles of the diffuse emission across the IRDC, away from point sources, at 8 and 24~$\mu$m. The surface brightness drops from 80 to 60 MJy/sr at 8~$\mu$m and from 40 to 30 at 24~$\mu$m. Taking into account the presence of foreground diffuse emission, we infer the depth at 8 and 24~$\mu$m due to the IRDC has a lower limit of 0.29 which corresponds to $A_V \ge 6$ mag, using $A_8/A_{V}\sim0.05$ and $A_{24}/A_{V}\sim0.05$ \citep{Indebetouw2005,Chapman2009}. The lower and upper-limits of the extinction along the line of sight are significantly different but they bracket the value found for the best fit of the IR SED. We are therefore confident with the correction we use.

Within this model, we cannot constrain the nature of the central object since its contribution to the NIR SED is not significant. However, we put an upper-limit on $W_*$ the dilution factor of the star, assuming it does not contribute significantly to the 2MASS J band. We assume the central source, claimed to be a Be/B[e]/LBV by \citet{Wachter2010}, has the radius and temperature of a B star: $R_*/R_\odot = 2$ to 7, $T_* = 10000$ to 30000 K. We apply the extinction from the best fit. We estimate that a B star of 2$R_\odot$ (resp. 7$R_\odot$) does not contribute more than 10\% to the J band if it is farther away than 400 pc (resp. 2.4 kpc). However, to remain within the Galaxy, the distance to MB4121 cannot be more than a factor of a few larger that those estimates. At a distance of 400 pc (resp. 2.4 kpc), the MIPS 24~$\mu$m shell is about 0.06 pc (resp. 0.36 pc) wide. For comparison, the bright MIPS 24~$\mu$m shell of the LBV G79.26+0.46, at a distance of 1.7 kpc, is about 0.85 pc wide \citep{Jimenez2010}.

\paragraph{MIR gas lines}

The IRS spectrum of MB4121 is rich in iron lines, especially toward the central source. The spectrum of the outer shell shows a weak [\ion{Fe}{2}] 25.99~$\mu$m line, and questionably the [\ion{Fe}{2}] 24.52~$\mu$m line. The gas line fluxes of the lines detected toward the central source and corrected for extinction are given in Table \ref{tab:flux4121a}. These lines are not resolved with IRS ($R\sim600$). As we show in the previous section, the emission toward the central source as seen in the 2MASS, IRAC and MIPS 24~$\mu$m images, is not dominated by the star but by an inner envelope of hot dust. We assume the iron lines arise from that envelope too.

\begin{figure*}[!t]
\centering
\subfigure[]
	{\label{fig:4121_iron1}
	\includegraphics[angle=90,width=.4\linewidth]{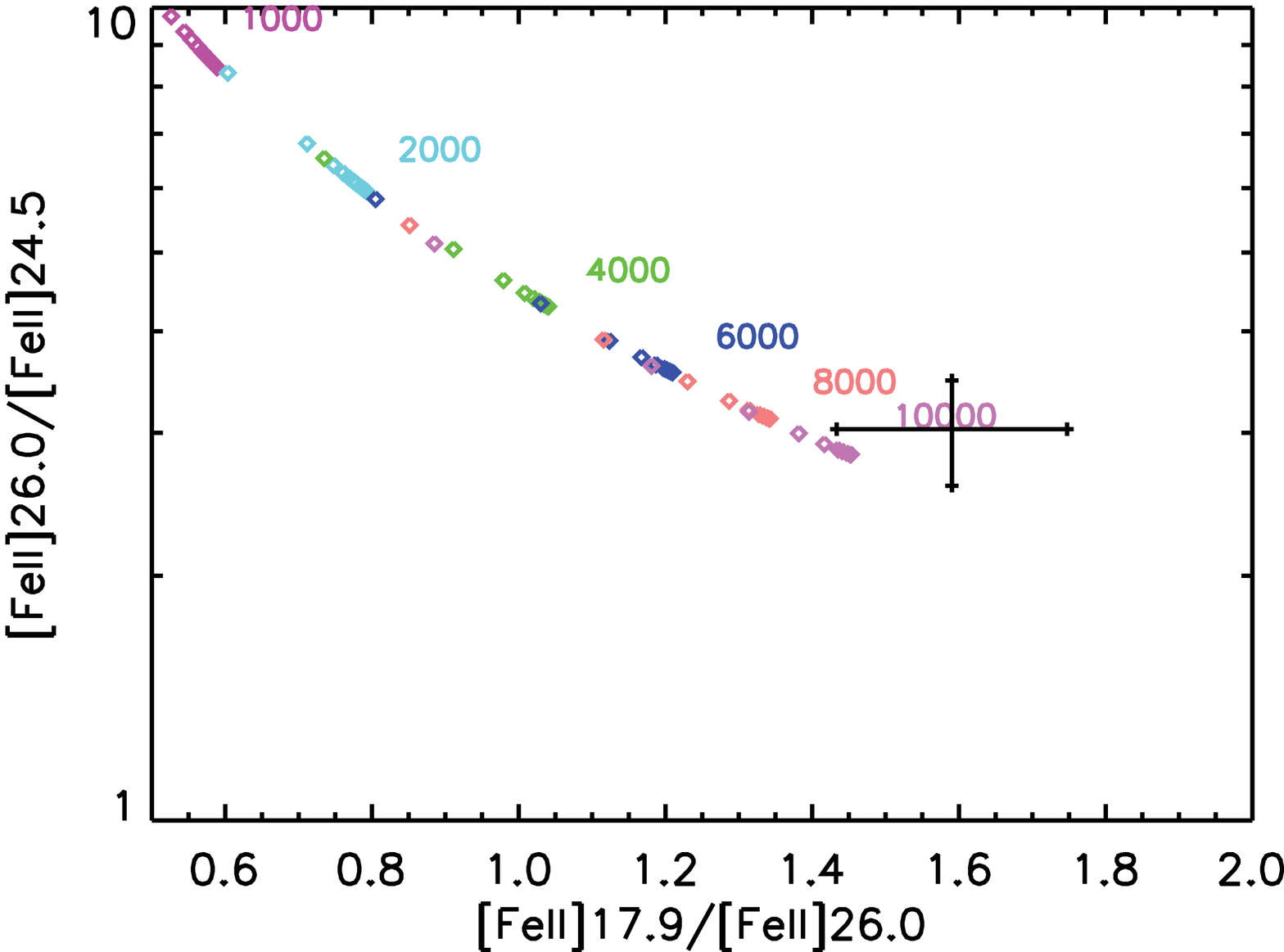}}
\subfigure[]
	{\label{fig:4121_iron2}
	\includegraphics[angle=90,width=.4\linewidth]{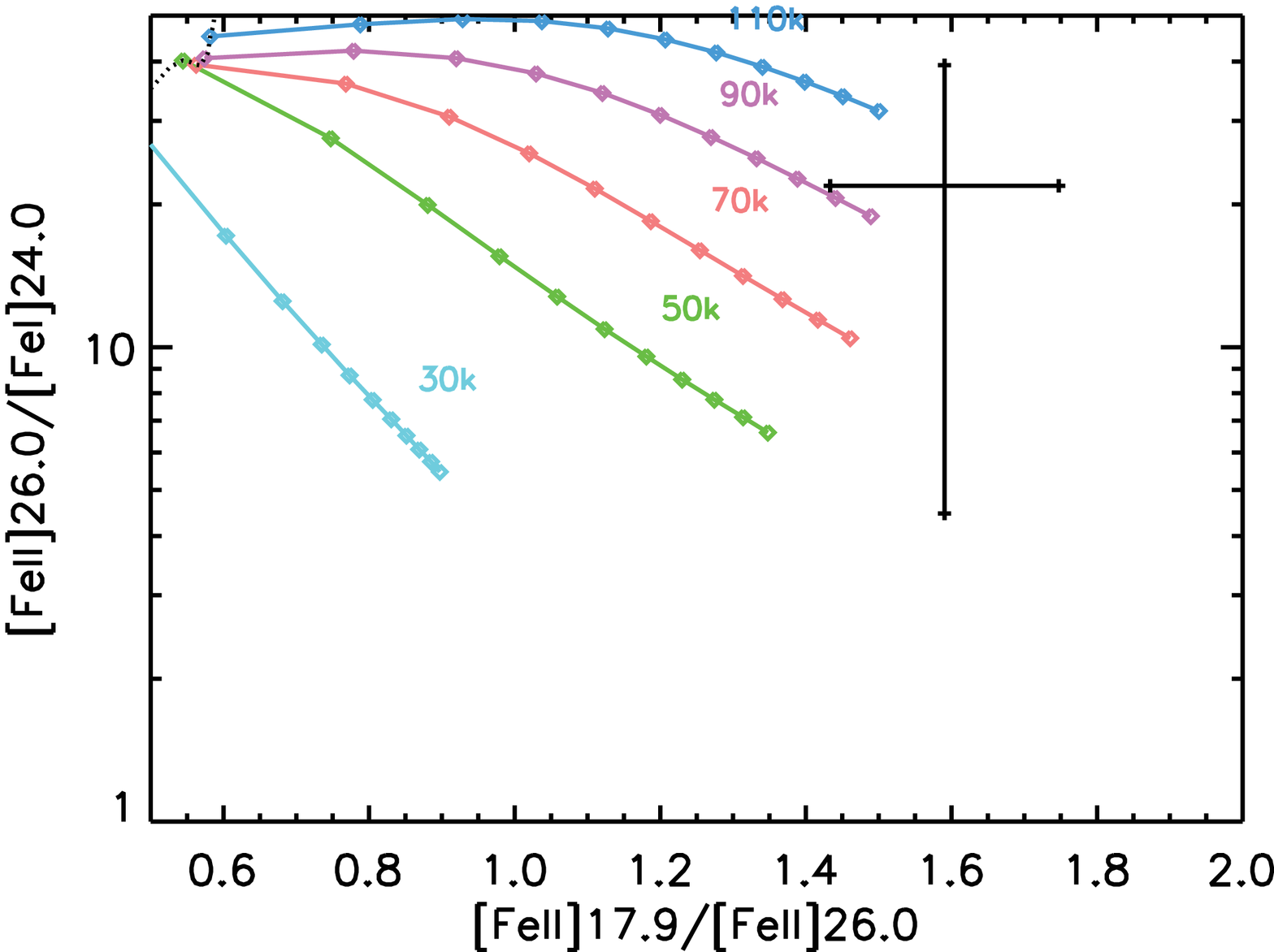}}
\caption{Diagrams of (a) the [\ion{Fe}{2}]~26.0/[\ion{Fe}{2}]~24.5~$\mu$m line ratio and (b) the [\ion{Fe}{2}]~26.0/[\ion{Fe}{1}]~24.0~$\mu$m line ratio both as a function of the [\ion{Fe}{2}]~17.9/[\ion{Fe}{2}]~26.0~$\mu$m line ratio, in a dust-free model. The color code correspons to (a) the given density (in cm$^{-3}$) or (b) the given temperature. All temperatures of the inner source are represented by a diamond in (a). Only densities between 6000 and 11000 cm$^{-3}$, from left to right, are represented by a diamond in (b). The cross indicates the observed iron lines ratios after correcting for the extinction.}
\label{fig:4121_iron}
\end{figure*}

\begin{figure*}[!t]
\centering
\subfigure[]
	{\label{fig:4121_iron1b}
	\includegraphics[angle=90,width=.4\linewidth]{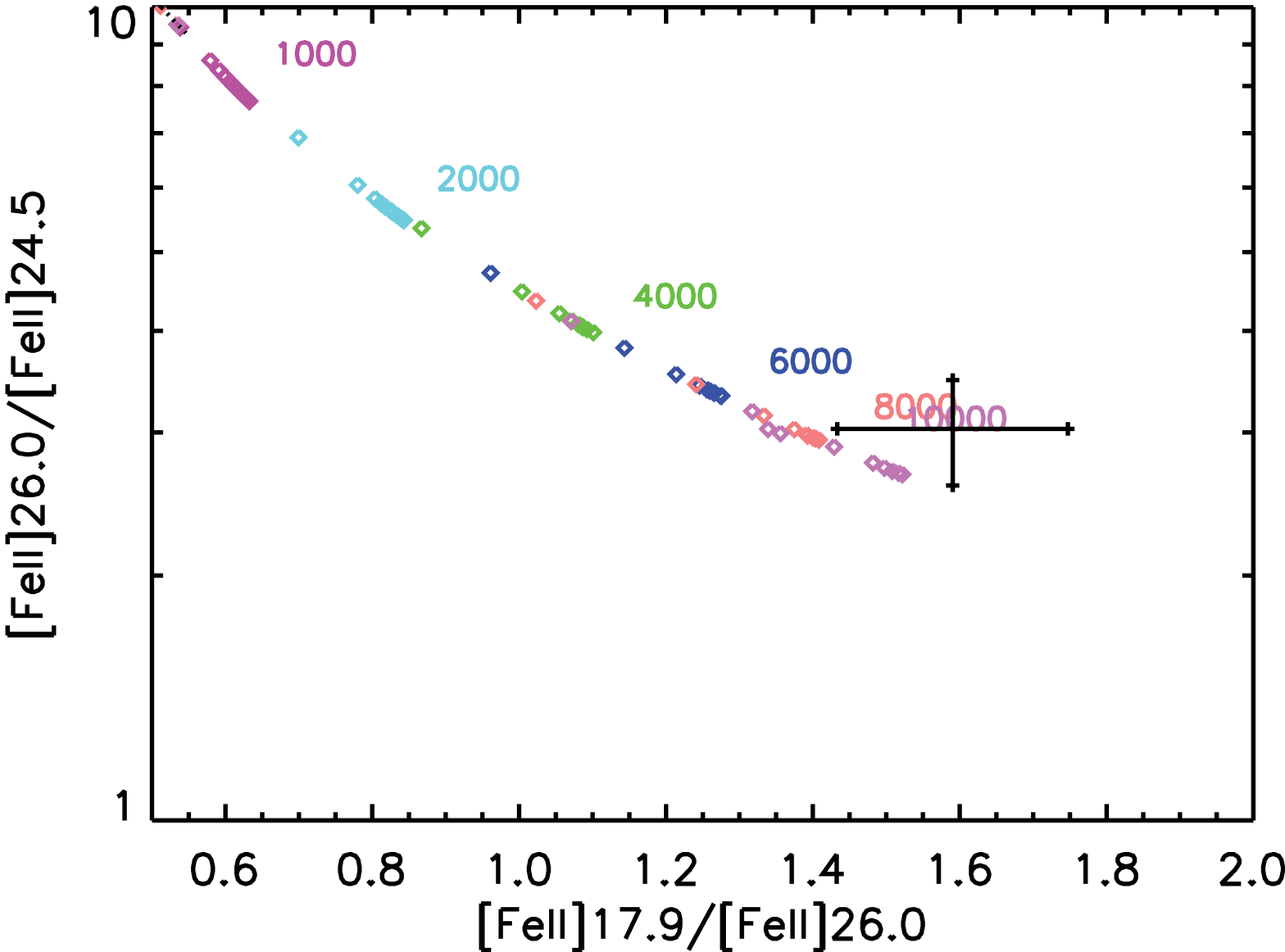}}
\subfigure[]
	{\label{fig:4121_iron2b}
	\includegraphics[angle=90,width=.4\linewidth]{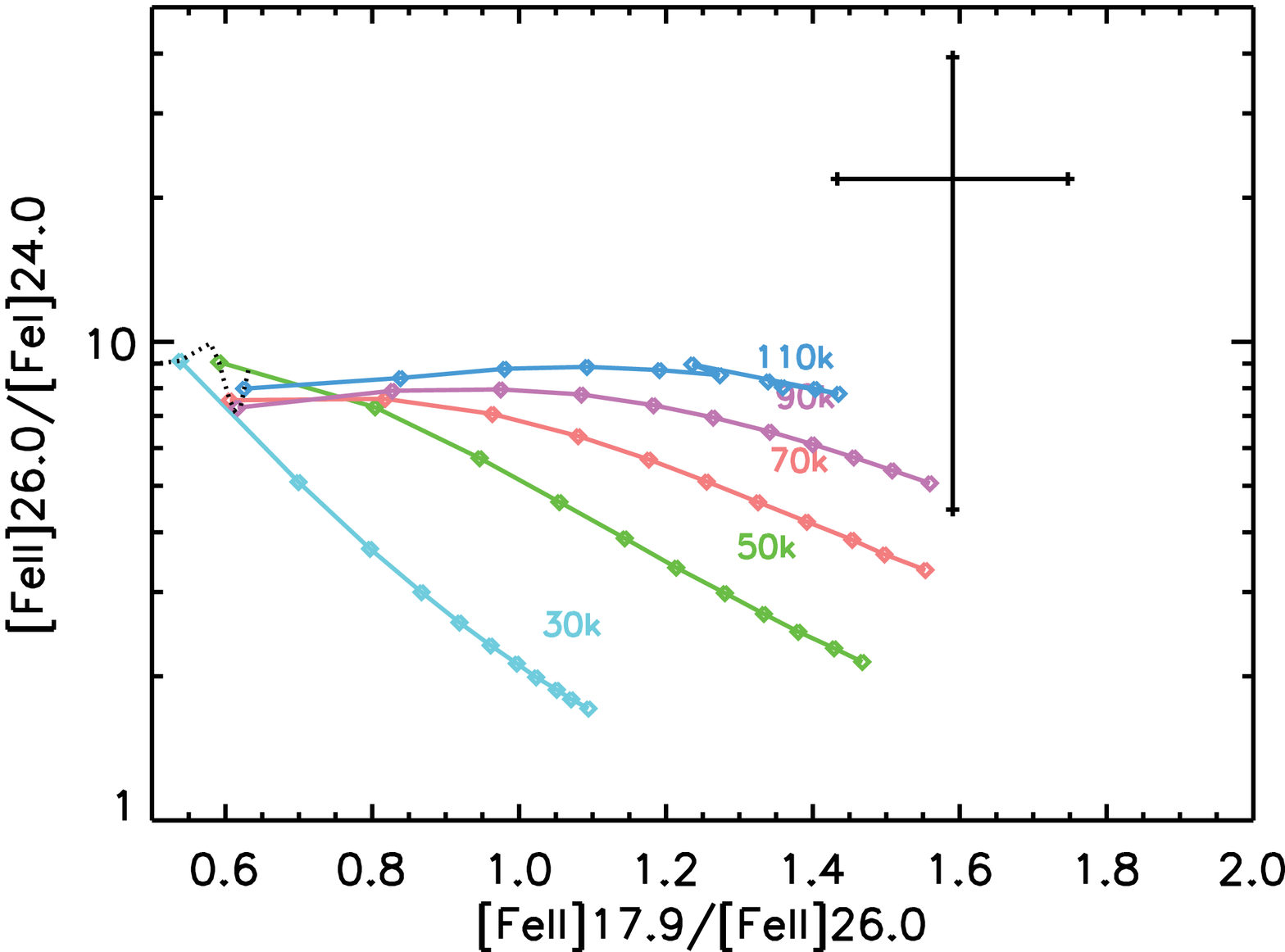}}
\caption{Same as previous figure but with dusty models.}
\label{fig:4121_ironb}
\end{figure*}

We first assume the iron lines fluxes can be explained within photoionization models. We use the results from the MAPPINGS~III model grid and compare to the observed brightest iron lines ([\ion{Fe}{2}]~17.94, 24.52 and 25.99~$\mu$m) to constrain the temperature of the central source and the density of the inner envelope. Figure \ref{fig:4121_iron1} shows the [\ion{Fe}{2}]~26.0/[\ion{Fe}{2}]~24.5 line ratio as a function of the [\ion{Fe}{2}]~17.9/[\ion{Fe}{2}]~26.0 line ratio for various gas densities and inner source temperatures. These iron lines, that we measured with a significant signal-to-noise ratio, only lead to a rough estimate of the gas density as they all are [\ion{Fe}{2}] lines. They indicate that, within the range of density and temperature that we explore, the gas density from where the [\ion{Fe}{2}] lines arise has to be at least $8\times10^3\ \rm{cm^{-3}}$. The addition of dust in the model does not significantly change this diagram (see Figure \ref{fig:4121_iron1b}). To constrain the inner source temperature, we use iron lines at different ionization level. In the IRS spectrum of MB4121, the only such line is that of [\ion{Fe}{1}] 24.04~$\mu$m whose uncertainty is significantly larger than that of the [\ion{Fe}{2}] lines. Figure \ref{fig:4121_iron2} shows the [\ion{Fe}{2}]~26.0/[\ion{Fe}{1}]~24.0 line ratio as a function of the [\ion{Fe}{2}]~17.9/[\ion{Fe}{2}]~26.0 line ratio for the same range of densities and temperature as in Figure \ref{fig:4121_iron1}. The best match is for a 90000 K central source but any temperature from 50000 to 110000 K is within 1-$\sigma$ of the observed ratios. The addition of dust to the model changes significantly this diagram. For a given [\ion{Fe}{2}]~26.0/[\ion{Fe}{1}]~24.0 line ratio, the required central source temperature is significantly higher when dust is present. Indeed, the photon energies at work to ionize the iron atom into Fe+ cations are between 7.9 and 16.2~eV (see Table \ref{tab:ip}). The dust is competing with the iron at those energy level, which corresponds to wavelenghts of 80 to 160 nm, as dust extinction is the highest in the UV and far-UV \citep[see e.g.][and reference therein]{Compiegne2011}. In MB4121, this competition between the extinction by the dust and the ionization of the iron atoms only occurs if the dust is not further away from the star than the iron atoms. In this case, the observed [\ion{Fe}{2}]~26.0/[\ion{Fe}{1}]~24.0 line ratio is in agreement with any temperature above 75000 K. Whether the dust is competing with the iron atom or not, the minimum required temperatures is about 50000 K which can hardly be accounted for by a Be/B[e]/LBV, as Be and B[e] stars have an effective temperature analog to those of B stars \citep[e.g.][]{Lamers1998} and LBV temperature is within the range 8000 to 40000 K \citep{Smith2004}. Within the limits we explore with our model, it thus seems difficult to confirm the Be/B[e]/LBV nature of MB4121 suggested by \citet{Wachter2010} unless it is an extraordinary hot source.

\begin{figure*}[!t]
\centering
\subfigure[]
	{\label{fig:4121_allen1}
	\includegraphics[angle=90,width=.4\linewidth]{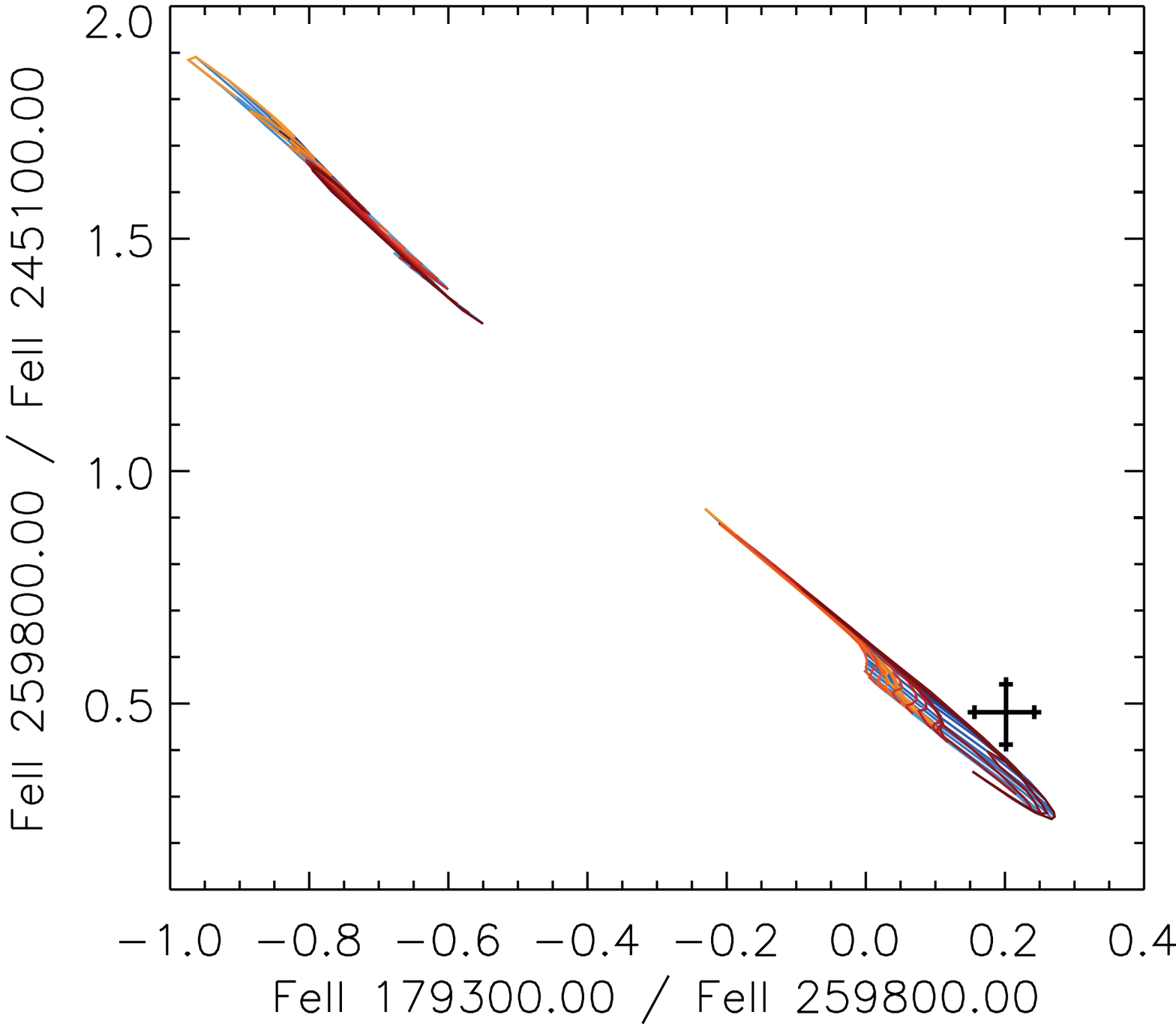}}
\subfigure[]
	{\label{fig:4121_allen2}
	\includegraphics[angle=90,width=.4\linewidth]{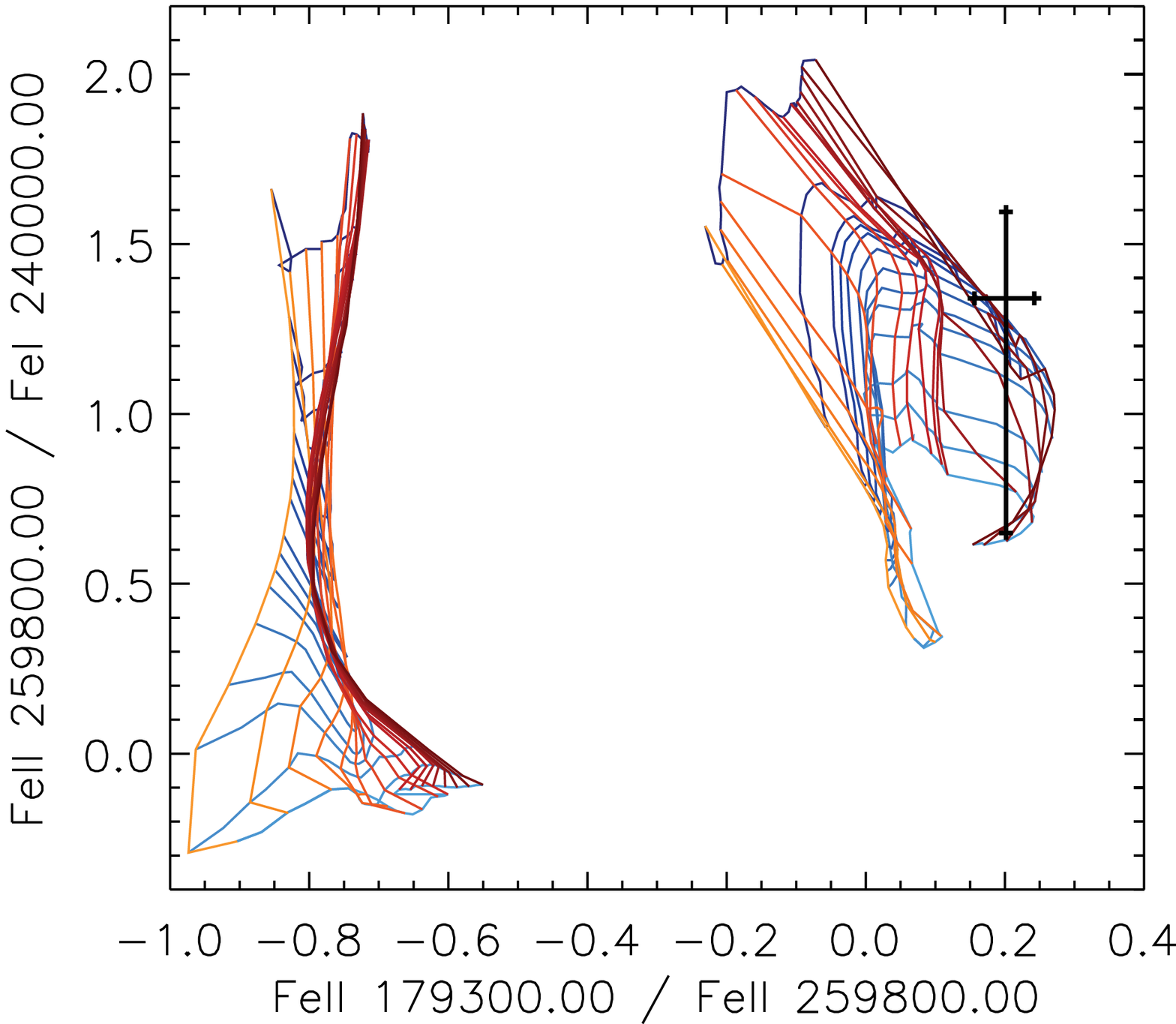}}
\caption{Diagrams of (a) the [\ion{Fe}{2}]~26.0/[\ion{Fe}{2}]~24.5~$\mu$m line ratio and (b) the [\ion{Fe}{2}]~26.0/[\ion{Fe}{1}]~24.0~$\mu$m line ratio both as a function of the [\ion{Fe}{2}]~17.9/[\ion{Fe}{2}]~26.0~$\mu$m line ratio from the MAPPINGS~III Library of Fast Radiative Shock Models \citep{Allen2008}. Axis scales are logarithmic. The curves on the left of both diagrams are for a density of 0.1 cm$^{-3}$ while the curves on the right are for a density of 1000 cm$^{-3}$. The red curves are for a given velocity, from 200 to 1000 km.s$^{-1}$, while the blue curves are for a given magnetic parameter $B/n^{1/2}$, from 0.0003 to 30 $\mu$G.cm$^{3/2}$.}
\label{fig:4121_allen}
\end{figure*}

We then attempt to interpret the observed iron line ratios with fast radiative shocks, using the models from \citet{Allen2008}. A shock is highly plausible in the vicinity of a Be/B[e]/LBV star as it is likely to undergo mass loss and/or have strong winds. Figure \ref{fig:4121_allen} shows the same line ratios as before for models of shock+precursor. We use the solar abundance models at two densities only, 0.1 and 1000 cm$^{-3}$, for clarity purposes. In both diagrams, the models are shown for a shock velocity between 200 and 1000 km.s$^{-1}$ and for a magnetic parameter $B/n^{1/2}$ between 0.0003 and 30 $\mu$G.cm$^{3/2}$. Both diagrams clearly show that the low density models cannot account for the observed ratios, whatever the shock velocity and the magnetic parameter. Indeed, only the highest density models from \citet{Allen2008} match the observed line ratios. In both diagrams, the modeled curves at 100 cm$^{-3}$, not shown here, are distant enough from those at 1000 cm$^{-3}$ that we can rule them out. Therefore we conclude that, if the iron lines are produced by a fast radiative shock, the density in the region from which the iron lines arise is 1000 cm$^{-3}$ within a factor of a few. Without better constraint on the density, we do not discuss furthermore these diagrams, which do not enable us to disentangle the different parameters. Indeed, both the shock velocity and magnetic parameter required to explain the observed ratios are highly dependent on the density. However, they allow us to find a set of parameters that would agree with the observations: a radiative shock of a few 100~km.s$^{-1}$ in a density of $\sim$1000~cm$^{-3}$. Since none of the lines are resolved with IRS, the velocity of a shock in MB4121 might not be higher than 500~km.s$^{-1}$.

We note here that the iron lines arise from an envelope near the central source which is unresolved in the MIPS 24~$\mu$m image. Therefore, its angular size is at most 6\arcsec. At a distance of 400 pc, the minimum distance determined in section \ref{lab:4121_fullfit}, it corresponds to 2400 AU or $3.6\times10^{16}\ \rm{cm}$. With a density of 1000 cm$^{-3}$, the total column density of the inner envelope is about a few $10^{19}\ \rm{cm^{-2}}$ which is equivalent to a visual extinction of about 0.02 mag. If MB4121 is farther away (e.g. 4 kpc) and the density a factor of a few higher than 1000 cm$^{-3}$, the extinction due to the inner envelope is at most 1 mag.

\paragraph{The outer shell}

The IRS spectra of MB4121 both show a continuum at wavelengths longer than $\sim 20 \mu$m. This continuum is detected with very similar intensities and spectral shape toward both position (see Figure \ref{fig:4121_irs}) which proves it originates in the outer shell of MB4121 detected in the MIPS 24~$\mu$m image. In this section, we discuss the properties of the dust responsible for the continuum emission and compare it with the flux measured in the MIPS 70~$\mu$m observations.

\begin{figure}[!t]
\centering
\subfigure[]
	{\label{}
	\includegraphics[angle=90,width=\linewidth]{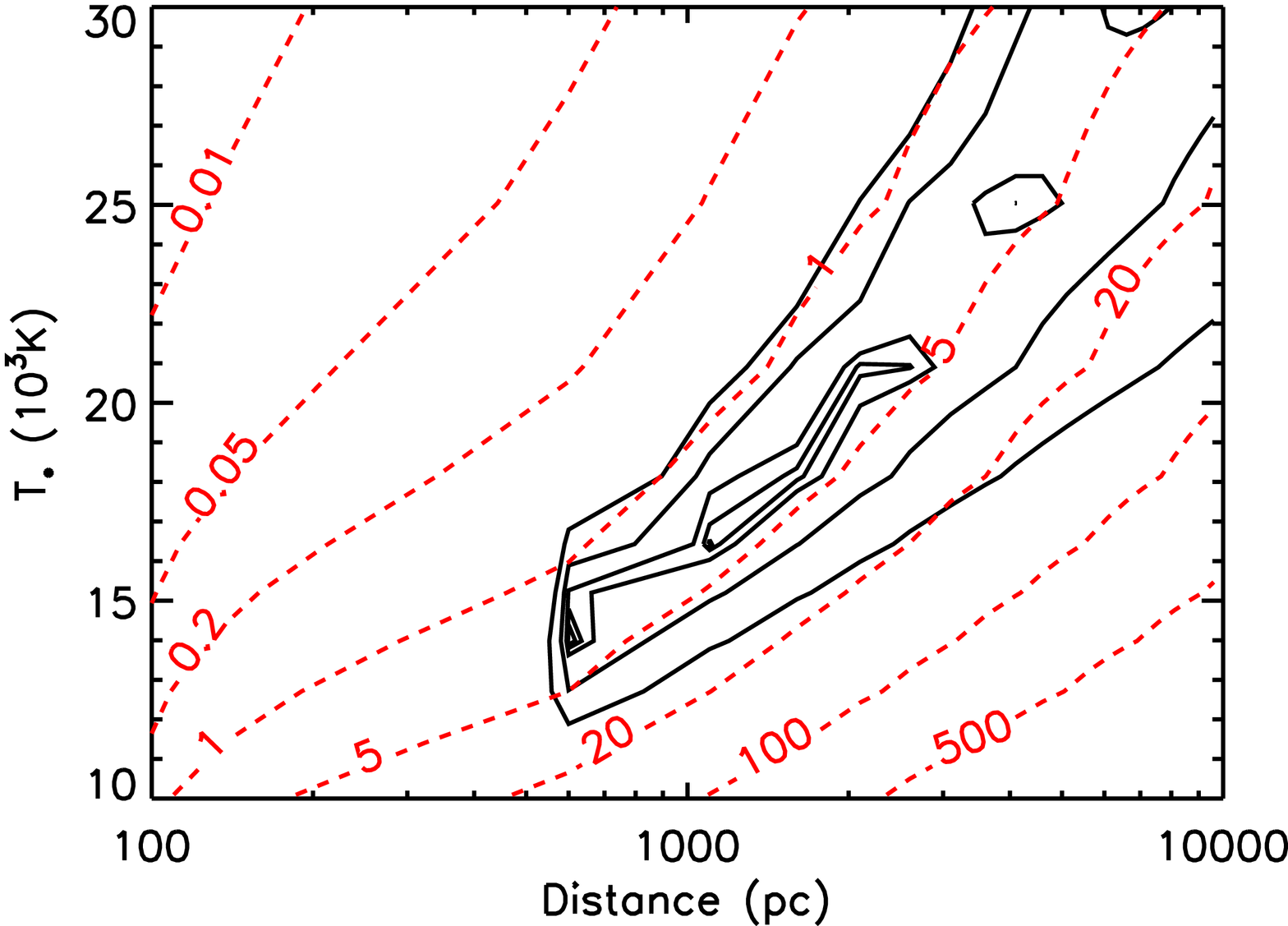}}
\subfigure[]
	{\label{}
	\includegraphics[angle=90,width=\linewidth]{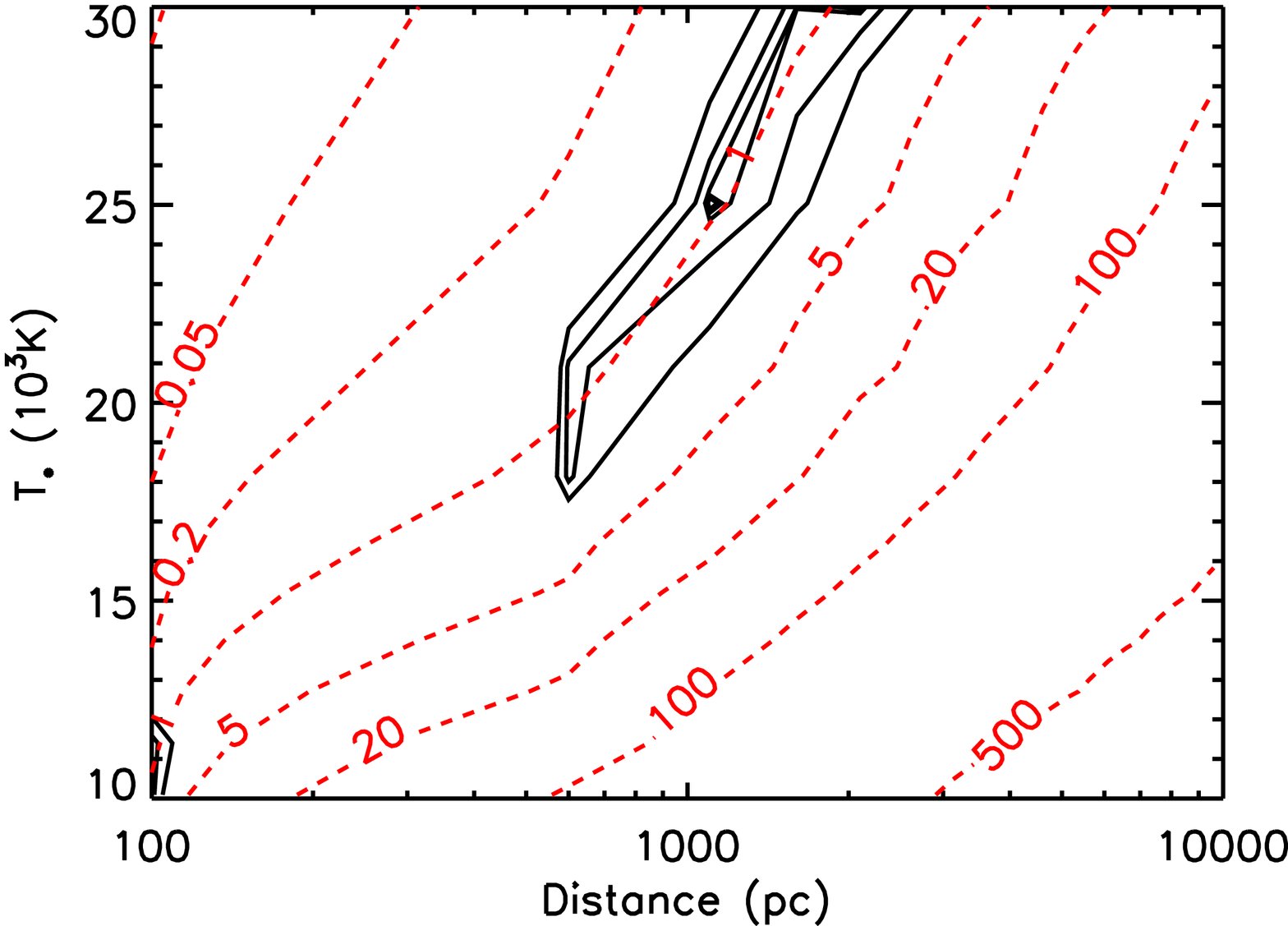}}
\subfigure[]
	{\label{}
	\includegraphics[angle=90,width=\linewidth]{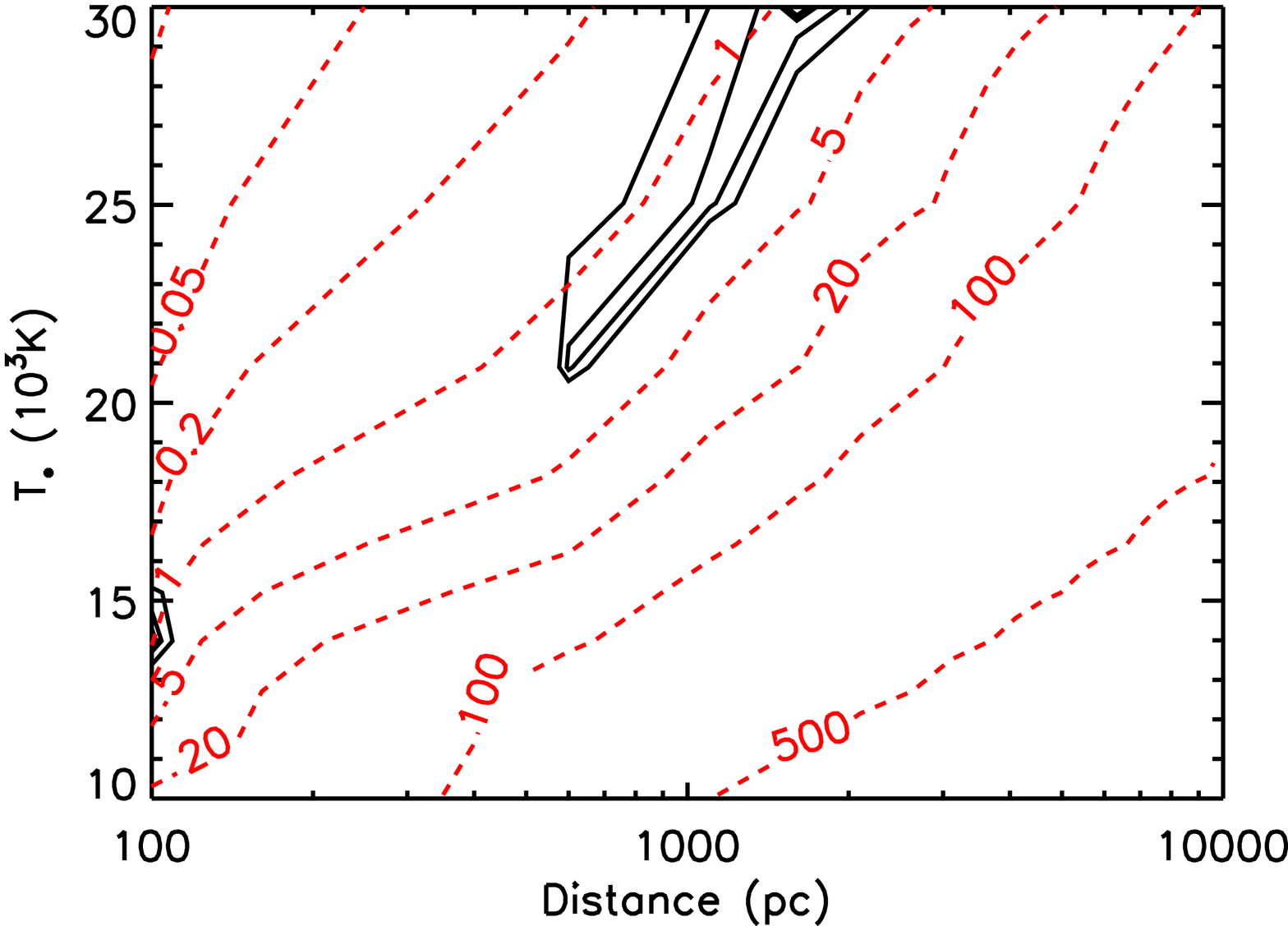}}
\caption{$\chi^2$ contours (solid black lines)for the fit of the long wavelength IRS continuum with the dust model and infered dust column density contours (red dashed lines) in $\mu$g.cm$^{-2}$. The dust size distribution is (a) SamC only, (b) LamC only and (c) aSil only.}
\label{fig:4121_shell}
\end{figure}

We use the IRS spectrum toward the central source of MB4121. The extinction correction performed on this spectrum (see Figure \ref{fig:4121_best_fit} and Section \ref{lab:4121_fullfit}) shows that a graybody ($\nu^2 \times B_\nu(T)$ function) with a temperature of 74~K best fits the continuum. We use the dust model from \citet{Compiegne2011} to estimate the dust mass in the MIPS 24~$\mu$m shell and the mass loss rate of the central source. Since we do not have any constraint on the dust size distribution in the shell, we consider it is dominated by either of the three following dust components: small amorphous carbon grains (SamC), large amorphous carbon grains (LamC) or amorphous silicates (aSil) \citep[see][for details about the component properties]{Compiegne2011}. We use a 20\arcsec\ radius for the MIPS 24~$\mu$m outer shell. The physical radius of the MIPS 24~$\mu$m shell thus is about $0.13\times d_*({\rm kpc}) \rm{pc}$ (i.e. 0.06 and 0.36 pc for a 2 and 7 $R_\odot$ respectively, using the distances from section \ref{lab:4121_fullfit}). The age of the shell then is $t = 1240 \times d_*({\rm kpc})/(v_{wind}/{\rm 100\ km.s^{-1}})$ years (i.e. 500 and 3000 years respectively). To model the dust emission we generate the radiation field for a B star with a radius between 2 and 7 $R_\odot$ and a temperature set accordingly between 10000 and 30000 K. We use a distance to MB4121 between 100 and 5000 pc. We apply to the model spectra the extinction curve deduced from the whole SED fit. We look for the column density of dust required to best match the IRS observations for wavelength $\lambda > 20 \mu$m. We find a column density of $2-3\times10^{-6}\ \rm{g.cm^{-2}}$ if we use SamC only and about $1\times10^{-6}\ \rm{g.cm^{-2}}$ if we use LamC or aSil only (see Figure \ref{fig:4121_shell}). We then estimate the total mass of dust within the MIPS 24~$\mu$m shell to be $5.7\times10^{-5}\times d_*({\rm kpc}) \ \rm{M_\odot}$ if the dust is mainly composed of large grains and a factor 2 to 3 larger if the dust is essentialy made of smaller grains. The total mass of the shell depends on the dust-to-gas mass ratio $x_d$ in the shell that is 1\% in the solar neighborhood but can be as small as 0.25\% \citep[e.g.][]{Sahai2000}. We finally estimate the mass loss rate of MB4121 to be $5.7 \times 10^{-5} \times d_*({\rm kpc}) \times (0.01/x_d) \times (v_{wind}/{\rm 100 km.s^{-1}})\ \rm{M_\odot/yr}$ which is in agreement with what is expected in LBVs. We finally compare the MIPS 70~$\mu$m flux measured at the position of the IRS-LH slit with that extrapolated from the model. The MIPS 70~$\mu$m measured within the IRS-LH slit toward MB4121 is 2.0$\pm$0.3 Jy, taking into account the uncertainty for variations from one ``nodding'' position to another and the photometric uncertainty of the MIPSGAL 70~$\mu$m observation \citep[15\%, ][]{Carey2009}. The extrapolated MIPS 70~$\mu$m flux from the dust model fit with aSil only leads to an extrapolated MIPS 70~$\mu$m flux of 2.4$\pm$0.2 Jy toward the central source in agreement with the observed flux. If we use SamC or LamC, the extrapolation is more uncertain: 3.0$\pm$0.4 Jy and 4.1$\pm$0.8 Jy, slightly higher than the observed value.

All those estimates are given for the distances we determine assuming the central star's contribution to the 2MASS J band. If the real distance is a factor of a few larger than our estimates, the timescale to make the shell is lowered by that same factor and the dust mass is increased by the square of this factor. As a result, the mass loss could be higher by a factor of a few. However, the distance to MB4121 cannot be significantly higher than a few kpc as it surely is a Galactic object. Indeed, the modeling of the outer shell continuum emission brings an additional constraint on the distance to MB4121. As Figure \ref{fig:4121_shell} shows, assuming the temperature of the central source is not higher than 30000 K, we can only find a good agreement with the observations if the distance to MB4121 is less than $\sim8$ kpc. If the dust in the MIPS 24~$\mu$m shell is dominated by large grains (LamC or aSil) then the distance to MB4121 has to be lower than $\sim2$ kpc. The modeling presented here has its limitations. For instance, it does only take into account radiative heating of the dust due to the central source and not collisional heating, which would increase the upper-limit on the distance to MB4121.

The combination of 2MASS, IRAC and IRS data allows us to model with a limited number of components the entire IR SED of MB4121. Contrary to what we model in MB3957, the central source of MB4121 is not contributing significantly to the IR emission. Therefore, we hardly constrain the distance to MB4121 as well as the temperature of its central source. The entire IR SED of MB4121 can be explained with three dust components from 75 K in the outer shell to 1250 K in the innermost region. Whether there are three distinct components or a continuous distribution of temperature is beyond the scope of this paper. The modeling of the IRS spectrum with the dust model of \citet{Compiegne2011} is in agreement with the Be/B[e]/LBV nature of MB4121, as suggested by \citet{Wachter2010}, at Galactic distances. We note however that the 2MASS, IRAC and IRS data were obtained a few years apart from each other which may be long enough to observe flux variation for LBVs \citet{Smith2004}. However, within the limits of our modelling, no variability is required.

\section{Conclusion}
\label{sec:ccl}

We present the results from our high-resolution Spitzer/IRS mid-IR spectroscopic observations of four MBs among the more than 400 detected in the MIPSGAL 24~$\mu$m survey of the Galactic plane. We combine these observations with previous 2MASS and IRAC observations, when available. We model the mid-IR gas lines detected and the dust emission to constrain the physical properties of the shell and their central source and therefore determine the nature of these objects. Details on each object follow:

\begin{itemize}
\item two MBs (MB4001 and MB4006) exhibit a dust-poor IRS spectrum dominated by highly-ionized gas lines (e.g. [\ion{O}{4}]~25.9, [\ion{Ne}{5}]~14.3 and 24.3~$\mu$m). Comparison with spectra of known planetary nebulae and modeling of the gas line ratios with MAPPINGS~III leads to the conclusion that these two MBs are planetary nebulae with a very hot ($\gtrsim 200,000$K) central white dwarf. We relate the almost total absence of dust emission features to the evolved stage in which the MBs might be.
\item MB3957 exhibits a dust-rich IRS spectrum and has a central source detected in the 2MASS and IRAC observations. The IRS spectrum is characterized by the presence of a few gas lines, among which the [\ion{Fe}{3}]~22.9~$\mu$m, [\ion{S}{4}]~10.5~$\mu$m and [\ion{Ne}{3}]~15.6~$\mu$m lines which rules out the shock excitation mechanism and leads to a strong constraint on the inner source temperature ($\sim 60000$K). Only a white dwarf or a Wolf-Rayet star can match such a temperature. The modeling of the IR SED of MB3957, from 1 to 40~$\mu$m, enables us to constrain the dilution factor of the central source and therefore conclude only a Wolf-Rayet star is a likely solution. The modeling also shows that two dust components (at 167 and 1750K), about 14 magnitudes of visual extinction and a distance of 7.5 kpc are necessary to interpret the IRS spectrum of MB3957. The MIPS 24~$\mu$m dusty shell comprises $\sim10^{-3}\ \rm{M_\odot}$. Assuming a wind velocity of 1000~km/s and a dust-to-gas mass ratio of 1\%, it implies the outer shell of MB3957 is $\sim425$ years old and that the average mass loss is $\sim10^{-6}\ \rm{M_\odot/yr}$.
\item MB4121 exhibits a dust-rich spectrum and has a central source detected in all IR images from 2MASS to MIPS 24~$\mu$m. Our multiple mid-IR spectra toward this object reveal many iron lines that arise in the central source and absorption features due to a infrared dark cloud along the line of sight. It was suggested as a Be/B[e]/LBV thanks to near-IR observations of its central source. We model the IR emission toward the central source, from 1 to 40~$\mu$m with three dust components, from 74 to 1250 K and 19 magnitudes of visual extinction. The central source contribution to the emission is required to be small. Therefore, we infer MB4121 is at least 400 pc away from us. The extinction corrected gas lines cannot be fitted by photoionization model but agree with fast radiative shock models for density of $\sim1000\ \rm{cm^{-3}}$. The modelling of the MIPS 24~$\mu$m outer dust shell results in a total mass of a few $10^{-3}\ \rm{M_\odot}$. It also implies MB4121 cannot be farther away than 8 kpc. Assuming a wind velocity of 100~km/s and a dust-to-gas mass ratio of 1\%, it implies the outer shell of MB4121 is $\sim1240$ years old and that the average mass loss is a few $10^{-5}\ \rm{M_\odot/yr}$. The distance to MB4121 is uncertain within more than an order of magnitude, so are these estimates. However, they all fit with the Be/B[e]/LBV interpretation.
\end{itemize}

With only four MBs observed in this program, we cannot draw any conclusions on the entire set of 428 objects detected in the Galactic plane. However, the combination of ongoing observational programs, from the optical to the radio, may soon lead to the first statistical results on the MIPSGAL bubbles and the population of evolved massive stars in our Galaxy.

\bibliographystyle{aa} 
\bibliography{../../../nicolasflagey} 

\end{document}